\documentclass[twocolumn]{aastex62}  % ApJ look-alike
\newcommand\apjcls{1}
\newcommand\aastexcls{2}
\newcommand\othercls{3}

\newcommand\papercls{\aastexcls}
\if\papercls \apjcls
\usepackage{apjfonts}
\else\if\papercls \othercls
\usepackage{epsfig}
\usepackage{margin}
\usepackage{times}
\fi\fi
\usepackage{ifthen}
\usepackage{natbib}
\usepackage{amssymb, amsmath}
\usepackage{appendix}
\usepackage{etoolbox}
\usepackage[T1]{fontenc}
\usepackage{paralist}

\if\papercls \apjcls
\newcommand\aas{\ref@jnl{AAS Meeting Abstracts}}% *** added by jh
\newcommand\dps{\ref@jnl{AAS/DPS Meeting Abstracts}}% *** added by jh
\newcommand\maps{\ref@jnl{MAPS}}% *** added by jh
\else\if\papercls \othercls
\usepackage{astjnlabbrev-jh}
\fi\fi

\if\papercls \aastexcls
\hypersetup{citecolor=blue, % color for \cite{...} links
            linkcolor=blue, % color for \ref{...} links
            menucolor=blue, % color for Acrobat menu buttons
            urlcolor=blue}  % color for \url{...} links
\else
\usepackage[%pdftex,      % hyper-references for pdflatex
bookmarks=true,           %%% generate bookmarks ...
bookmarksnumbered=true,   %%% ... with numbers
colorlinks=true,          % links are colored
citecolor=blue,           % color for \cite{...} links
linkcolor=blue,           % color for \ref{...} links
menucolor=blue,           % color for Acrobat menu buttons
urlcolor=blue,            % color for \url{...} links
linkbordercolor={0 0 1},  %%% blue frames around links
pdfborder={0 0 1},
frenchlinks=true]{hyperref}
\fi

\if\papercls \othercls

\else

\fi

\providecommand{\adsurl}[1]{\href{#1}{ADS}}

\makeatletter
\patchcmd{\NAT@citex}
  {\@citea\NAT@hyper@{%
     \NAT@nmfmt{\NAT@nm}%
     \hyper@natlinkbreak{\NAT@aysep\NAT@spacechar}{\@citeb\@extra@b@citeb}%
     \NAT@date}}
  {\@citea\NAT@nmfmt{\NAT@nm}%
   \NAT@aysep\NAT@spacechar\NAT@hyper@{\NAT@date}}{}{}

\patchcmd{\NAT@citex}
  {\@citea\NAT@hyper@{%
     \NAT@nmfmt{\NAT@nm}%
     \hyper@natlinkbreak{\NAT@spacechar\NAT@@open\if*#1*\else#1\NAT@spacechar\fi}%
       {\@citeb\@extra@b@citeb}%
     \NAT@date}}
  {\@citea\NAT@nmfmt{\NAT@nm}%
   \NAT@spacechar\NAT@@open\if*#1*\else#1\NAT@spacechar\fi\NAT@hyper@{\NAT@date}}
  {}{}
\makeatother

\makeatletter
\DeclareRobustCommand{\lowcase}[1]{\@lowcase#1\@nil}
\def\@lowcase#1\@nil{\if\relax#1\relax\else\MakeLowercase{#1}\fi}
\makeatother

\DeclareSymbolFont{UPM}{U}{eur}{m}{n}
\DeclareMathSymbol{\umu}{0}{UPM}{"16}
\let\oldumu=\umu
\renewcommand\umu{\ifmmode\oldumu\else\math{\oldumu}\fi}

\if\papercls \othercls

\else

\fi

\let\oldsim=\sim
\renewcommand\sim{\ifmmode\oldsim\else\math{\oldsim}\fi}
\let\oldpm=\pm
\renewcommand\pm{\ifmmode\oldpm\else\math{\oldpm}\fi}
\newcommand\by{\ifmmode\times\else\math{\times}\fi}
\newbox{\wdbox}
\renewcommand\c{\setbox\wdbox=\hbox{,}\hspace{\wd\wdbox}}
\renewcommand\i{\setbox\wdbox=\hbox{i}\hspace{\wd\wdbox}}

\newcount\timect
\newcount\hourct
\newcount\minct
\newcommand\now{\timect=\time \divide\timect by 60
         \hourct=\timect \multiply\hourct by 60
         \minct=\time \advance\minct by -\hourct
         \number\timect:\ifnum \minct < 10 0\fi\number\minct}

\catcode`@=11

\newcommand\comment[1]{}

\newcommand\commenton{\catcode`\%=14}

\renewcommand\math[1]{$#1$}
\newcommand\mathshifton{\catcode`\$=3}

\let\atab=&
\newcommand\atabon{\catcode`\&=4}

\let\oldmsp=\sp
\let\oldmsb=\sb
\def\sp#1{\ifmmode
           \oldmsp{#1}%
         \else\strut\raise.85ex\hbox{\scriptsize #1}\fi}
\def\sb#1{\ifmmode
           \oldmsb{#1}%
         \else\strut\raise-.54ex\hbox{\scriptsize #1}\fi}
\newbox\@sp
\newbox\@sb
\def\sbp#1#2{\ifmmode%
           \oldmsb{#1}\oldmsp{#2}%
         \else
           \setbox\@sb=\hbox{\sb{#1}}%
           \setbox\@sp=\hbox{\sp{#2}}%
           \rlap{\copy\@sb}\copy\@sp
           \ifdim \wd\@sb >\wd\@sp
             \hskip -\wd\@sp \hskip \wd\@sb
           \fi
        \fi}
\def\msp#1{\ifmmode
           \oldmsp{#1}
         \else \math{\oldmsp{#1}}\fi}
\def\msb#1{\ifmmode
           \oldmsb{#1}
         \else \math{\oldmsb{#1}}\fi}

\def\supon{\catcode`\^=7}

\def\subon{\catcode`\_=8}

\def\supsubon{\supon \subon}

\newcommand\actcharon{\catcode`\~=13}

\newcommand\paramon{\catcode`\#=6}

\comment{And now to turn us totally on and off...}

\newcommand\reservedcharson{ \commenton  \mathshifton  \atabon  \supsubon 
                             \actcharon  \paramon}

\catcode`@=12
\reservedcharson

\if\papercls \apjcls

\else

\fi

\newcommand\chisq{\ifmmode{\chi\sp{2}}\else\math{\chi\sp{2}}\fi}
\newcommand\redchisq{\ifmmode{ \chi\sp{2}\sb{\rm red}}
                    \else\math{\chi\sp{2}\sb{\rm red}}\fi}
\newcommand\Teq{\ifmmode{T\sb{\rm eq}}\else$T$\sb{eq}\fi}
\newcommand\Teff{\ifmmode{T\sb{\rm eff}}\else$T$\sb{eff}\fi}
\newcommand\mjup{\ifmmode{M\sb{\rm Jup}}\else$M$\sb{Jup}\fi}
\newcommand\rjup{\ifmmode{R\sb{\rm Jup}}\else$R$\sb{Jup}\fi}
\newcommand\msun{\ifmmode{M\sb{\odot}}\else$M\sb{\odot}$\fi}
\newcommand\rsun{\ifmmode{R\sb{\odot}}\else$R\sb{\odot}$\fi}
\newcommand\mearth{\ifmmode{M\sb{\oplus}}\else$M\sb{\oplus}$\fi}
\newcommand\rearth{\ifmmode{R\sb{\oplus}}\else$R\sb{\oplus}$\fi}
\usepackage{amsmath}
\usepackage{graphicx}

\usepackage{epsfig}
\usepackage{natbib}
\usepackage{wasysym}
\usepackage{lipsum}
\usepackage{mathrsfs}
\usepackage{float}
\usepackage{rotating}
\usepackage{graphicx}
\usepackage{epstopdf}
\usepackage{colortbl}
\usepackage{hyperref}
\newcommand{\angstrom}{\textup{\AA}}
\providecommand{\e}[1]{\ensuremath{\times 10^{#1}}}

\usepackage{color}

\definecolor{twitterblue}{RGB}{64,153,255}

\usepackage{hyperref}

\usepackage{filecontents}

\begin{document}

\shorttitle{Planet and Atmospheric Co-formation on TRAPPIST-1 Analogs}
\shortauthors{Chen, H., Clement, M. S., Wang, L. C., \& Gu, J. T.}

\title{{\bf  Born Dry or Born Wet? A Palette of Water Growth Histories in \\TRAPPIST-1 Analogs and Compact Planetary Systems}}

\author[0000-0003-1995-1351]{Howard Chen}

\affil{Department of Aerospace, Physics, and Space Sciences, Florida Institute of Technology, Melbourne, FL 32901}

\affil{Sellers Exoplanet Environments Collaboration (SEEC), NASA Goddard Space Flight Center, Greenbelt, MD 20771}

\affil{Consortium on Habitability and Atmospheres of M-dwarf Planets (CHAMPs), Laurel,  MD 20723}
\author[0000-0001-8933-6878]{Matthew S Clement}
\affil{Consortium on Habitability and Atmospheres of M-dwarf Planets (CHAMPs), Laurel,  MD 20723}

\affil{Applied Physics Laboratory, Johns Hopkins University, Laurel, MD 20723}

\author[0000-0002-6379-3816]{Le ``Chris" Wang}

\affil{Department of Physics \& Astronomy, Johns Hopkins University, Baltimore, MD 21218}

\affil{Department of Astrophysical Sciences, Princeton University, Princeton, NJ 08544}

\author[0000-0002-6625-6346]{Jesse T. Gu}

\affil{Department of Earth and Planetary Sciences, Harvard University, Cambridge, MA 02138}

\begin{abstract}
It is still unclear whether exoplanets in compact multiplanet systems such as TRAPPIST-1 are able to accrete large quantities of volatiles, grow to sufficient mass, and maintain robust atmospheres and hydrospheres.  Previous estimates of water content in M-dwarf systems have largely relied on population synthesis or atmosphere–interior evolution models, often treating impacts and atmospheric loss in isolation. In this work, we couple impact delivery, impact erosion, and mantle–atmosphere exchange within a model that tracks volatile evolution through stochastic collision histories. By explicitly including both planetesimal accretion and the prolonged luminous pre-main-sequence phase of M dwarfs, we find lower water inventories for the inner TRAPPIST-1 analogs (b–e), spanning only $10^{-4}$–$10^{-2},M_{\oplus,\rm ocn}$ across a wide range of disk structures and impact scenarios.
By contrast, the outer planets (f–h analogs) frequently retain water inventories exceeding an Earth ocean mass. This systematic volatile gradient provides a physically motivated explanation for JWST’s nondetections of atmospheres on TRAPPIST-1 b and c, implying an origin rooted in formation conditions rather than in post-formation escape. Our results suggest that many rocky planets in compact M-dwarf systems may form already depleted in volatile compounds, fundamentally limiting their capacity to sustain atmospheres or surface oceans. More broadly, our multistage framework for volatile tracking can help interpret future observations of compact system and set more realistic initial conditions for exoplanet interior compositions and atmospheric models
\end{abstract}

\section{Introduction} 
\label{sec:intro}

TRAPPIST-1, the Teegarden star system, and Kepler-186 \citep{GillonEt2017NATURE, zechmeister2019,quintana2014} represent some of the most compelling multiplanetary systems for upcoming transmission spectroscopy campaigns and future direct imaging missions. These systems offer rare opportunities for comparative exoplanetology, as their constituent planets likely formed under similar initial conditions but may have diverged substantially through subsequent evolutionary processes. Key system properties, such as stellar age, metallicity, activity levels, and the diversity of planetary architectures, can be jointly constrained across multiple planets, helping to disentangle the intertwined effects of formation environment and dynamical evolution. However, many of these systems are extremely compact, with five or more planets orbiting within 0.1–0.5 AU. Their tight configurations imply formation in diminutive, low-mass protoplanetary disks, raising long-standing questions about the efficiency of water delivery and volatile retention during accretion \citep{lissauer2007, raymond2007, ciesla2015}. In particular, mechanisms proposed for volatile delivery in our own Solar System such as destabilization of icy planetesimals during giant planet formation \citep{Raymond+17}, inward migration of massive planets \citep{OBrienEt2014Icar}, and late-stage cometary delivery following the Nice Model instability \citep{joiret2023}  would likely be absent in these systems. For planets forming in compact, gas-depleted disks, it remains unclear whether inner protoplanets can accrete sufficient volatiles to sustain atmospheres and hydrospheres, or whether they are inherently dry.

Planets form within gas-rich, dust-laden disks surrounding young stars through a bottom-up process involving millions of high-energy collisions between planetesimals and growing protoplanets. This accretional phase unfolds over timescales of several million years \citep{Chambers2004EPSL, jacobson+morbi14}. During the late stages of this growth, episodes of strong radial mixing drive dynamical excitation and scattering of embryos and planetesimals across broad regions of the disk. These bodies are then incorporated into developing planets through a series of embryo–embryo and embryo–planetesimal mergers. Such giant impacts not only dictate the final mass and orbital architecture of terrestrial planets, but also play a pivotal role in setting their volatile inventories, that is, the abundance of light species such as water and CO$_2$ that typically reside in gaseous or vapor states \citep{venturini2020}.

The initial water endowment of a planet plays a critical role in shaping its long-term atmospheric evolution and potential for habitability. If a planet forms dry, it may fail to initiate key climate-regulating processes such as global silicate weathering, potentially locking it into an inhospitable runaway greenhouse state, an outcome proposed for Venus under dry initial conditions \citep{hamano2013}. Conversely, moderate amounts of surface and interior water are often considered essential for sustaining habitable environments, or at least those analogous to present-day Earth \citep[e.g.,][]{cockell2016}. On Earth, silicate weathering is largely driven by continental exposure, which in turn depends on a delicate balance between land and ocean coverage. With seawater comprising only 0.023\% of Earth’s total mass, sufficient land remains exposed to enable effective climate regulation. However, on ocean worlds with tens of times more surface water, silicate weathering may become inefficient or even suppressed entirely \citep{abbot2012,nakayama2019}. On the opposite extreme, extremely dry planets may lack plate tectonics: the key mechanism for recycling volatiles from the interior and sustaining secondary atmospheres over geological timescales \citep{komacek+16}. The absence of tectonic activity may also prevent a planet from transitioning from a CO$_2$-dominated atmosphere to an N$_2$-dominated one \citep{guo2025}.

Many prior studies assume that protoplanets inherit their volatile content from the surrounding nebular disk, with water-loss rates primarily governed by photoevaporation during the host star’s luminous X-ray and EUV phase. For instance, population synthesis models combined with hydrodynamic escape schemes predict a broad distribution of water contents for planets orbiting M-dwarfs \citep{miguel2020,tian2015,Luger+Barnes2015AsBio,kimura2022,childs2023}. Other investigations have explored the formation of secondary atmospheres through mantle degassing, thermal escape, and giant impacts, revealing that these processes play a central role in sculpting a planet’s atmospheric evolution \citep{krissansen2024,rogers2024,young2024,kurosaki2023,lock2024}.

Recent semianalytical models, such as those by \citet{muller2024}, suggest that outer TRAPPIST-1 planets could retain between 1\% and 10\% of Earth’s ocean mass in water. Complementary insights from molecular dynamics simulations (e.g., \citealt{solomatova2021}) point to the possibility of water-depleted but CO$_2$-rich initial atmospheres on rocky Earth-like planets. Together, these approaches highlight the range of plausible volatile inventories for low-mass exoplanets. Building on this foundation, our work emphasizes the additional role of planetesimal accretion, which is often not explicitly treated in these models. For rocky planets with masses $\lesssim 1,M_\oplus$, such as TRAPPIST-1 analogs, impact delivery from 1–500 km planetesimals may dominate volatile acquisition, underscoring the importance of coupling dynamical N-body histories with interior and atmospheric models.

While several studies have proposed the existence of H/He-dominated envelopes around temperate low-mass exoplanets \citep{owen2020,lammer2025}, rapid atmospheric boil-off in compact systems casts doubt on the long-term retention of such low mean molecular weight atmospheres. These processes may also erode or remove any initially accreted high mean molecular weight gases. Consequently, the dominant mechanism for volatile delivery, particularly for species such as H$_2$O, CO$_2$, and N$_2$, in these systems may be collisional in nature. This is analogous to the solar system, where volatile acquisition is believed to result largely from the accretion of chondritic material and icy bodies scattered inward from beyond the snowline \citep{chyba1994,schlichting+18,zahnle2006}. To date, however, such solid-body impact delivery pathways remain underexplored in the context of compact, tightly packed multiplanet systems.

During planetary accretion, volatiles are transferred from the nebular gas and solids into various planetary reservoirs. However,  subsequent processing stages, including core formation, magma ocean crystallization, and differentiation can incur losses and lead to depletion of even moderately volatile species \citep{dalou+17,gu2024}, noble gases \citep{marty+20}, and highly siderophile elements like carbon \citep{hirschmann+21}. A planet’s final volatile budget thus reflects both the composition and structure of its natal disk and the sequence of energetic events it experiences.

While these processes define the volatile inventories available to a young planet, the manner in which material is delivered is equally critical. Traditional models of planetesimal accretion often represent impacts as a constant, smooth flux derived from bulk dynamical histories (e.g., \citealt{Matsui+Abe1986,ZahnleEt1988Icar}), but this idealization overlooks the inherently stochastic nature of late-stage growth. N-body simulations reveal that accretion is a chaotic process \citep{Chambers2001Icar,clement2018}, where a single collision can dramatically alter a planet’s volatile budget. Yet many studies, particularly for systems around low-mass stars, still adopt perfect-merger assumptions, neglecting the mass loss and compositional changes caused by energetic impacts. Recent 1D and 3D modeling has shown that devolatilization depends sensitively on impact geometry, velocity, and the differentiation state of the colliding bodies \citep{Chen+Jacobson22,kegerreis2020,lock2024}. Incorporating these effects into models of Earth and Mars formation using chondritic impactors has successfully reproduced the volatile depletion patterns observed in the Solar System \citep{sakuraba+19,sakuraba2021}, underscoring the need to account for collision physics when interpreting planetary volatile budgets.

Because of this inherent stochasticity, the volatile retention or loss from any given collision depends on both external conditions (e.g., impact angle, velocity) and internal planetary structures (e.g., magma ocean depth, atmosphere mass) \citep{rubie+15,cambioni2021}. Thus, capturing the true diversity of outcomes requires models that track volatile transport across atmosphere, mantle, and core reservoirs on an event-by-event basis.

Volatile acquisition is further complicated by the coupling between dynamical evolution and compositional inheritance. Collision rates reflect the dynamical state of the gas disk, while each impactor’s composition depends on its formation location and migration history. These interdependencies demand an integrated approach linking disk chemistry, dynamical architecture, and interior evolution.

\begin{figure*}
\begin{center}
\includegraphics[width=1.8\columnwidth]{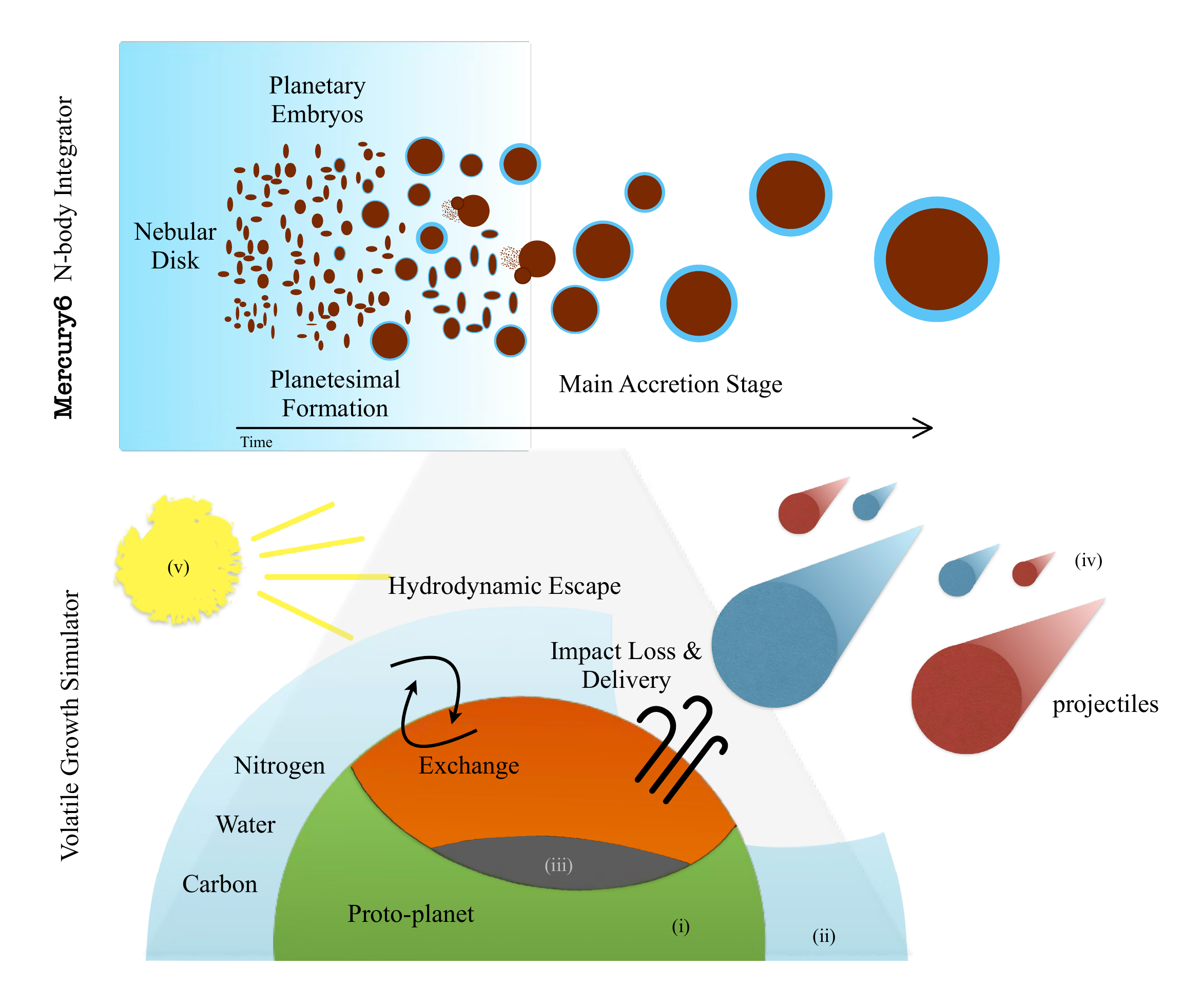}
\caption{\label{fig1_summary} Schematic of our modeling framework, illustrating the connection between N-body accretion outcomes and volatile growth simulations. Each panel highlights the primary physical processes represented at each stage of the model. Terrestrial planet formation begins with planetesimals and planetary embryos assumed to emerge from an initial gas+dust disk. Their subsequent growth is explicitly simulated using an N-body accretion model that incorporates processes such as collision fragmentation. The outputs from each collision event and accretion phase then serve as inputs to a self-consistent volatile growth model, which evolves key volatile species such as H$_2$O, C, and N, based on cosmochemically informed initial compositions. Major processes contributing to volatile delivery and loss include impact erosion and mantle degassing. Roman numerals denote model sub-components: mantle (i), atmosphere (ii), core (iii), impactors (iv), and stellar EUV activity (v). Components (i)–(iii) constitute the three primary reservoirs of the growing protoplanet. The schematic layout is conceptually inspired by \citet{gu2024}. } 
\end{center}
\end{figure*}

Here, we develop a novel framework that combines N-body accretion simulations with a volatile tracking model capable of following H$_2$O, CO$_2$, N$_2$ across three planetary reservoirs: atmosphere, mantle, and core (Figure~\ref{fig1_summary}). Our model incorporates impact-driven degassing/ingassing, atmospheric ablation, and core–mantle–atmosphere partitioning, using disk-derived volatile compositions as initial conditions. While prior studies have included some of these processes individually (e.g., \citealt{rubie+15}), few have integrated them into a time-resolved model that explicitly tracks planetary accretion histories.

We apply this model to TRAPPIST-1 analog systems to explore how volatile elements evolve across compact multiplanet environments. Our simulations generate time- and spatial-dependent predictions of water and carbon dioxide inventories in both surface and internal reservoirs. These results provide physically grounded initial conditions for interior and atmospheric models and may serve as future observables for next-generation telescopes.

\section{Numerical Models \& Methods}
\label{sec:methods}

\subsection{N-body Accretion Model}

We use the \verb|Mercury6| hybrid integrator \citep{chambers1999} with additional modifications to model the effects of collisional fragmentation \citep{chambers2013} to model planet accretion histories. The hybrid integrator is commonly used in studies of rocky planet accretion in both the solar system \citep{Chambers2001Icar} and exoplanet systems \citep{clement2022} and uses the Bulirsch-Stoer method to numerically integrate close encounters and a syplectic integrator when objects are separated at larger distances. Due to the stochastic nature of embryo accretion \citep{RaymondEt2006Icar}, we run a number of simulations for each tested set of initial configuration to broadly sample the statistical range of plausible system outcomes.

\subsubsection{Numerical Setup \& Integrations}

\paragraph{TRAPPIST-1 Systems}

Our TRAPPIST-1 N-body simulations are derived from a larger suite of models reported in \citet{Clement2025}.  Our simulations are designed to commence around the time that TRAPPIST-1's nebular gas disk has mostly dissipated, and the bulk of planet growth and orbital migration has already concluded.  Each model places a system of 30 large embryos in a a disk of 1,000 smaller planetesimals.  Both the embryo and planetesimal components of the disk carry equal total masses (between 3 and 4 Earth masses depending on the simulation set), as is common in previous studies of both the solar system and exoplanets \citep[e.g.][]{raymond2009building}.  All models employ inner and outer disk boundaries of 0.01 and 0.1 au, respectively, and assign orbital semi-major axes to bodies such that the disk's surface density profile falls off as $r^{-\alpha}$ (we test values of $\alpha=$ 1.0, 1.5 and 2.5).  Our inner disk edge is loosely motivated by the location of TRAPPIST-1's magnetic truncation radius at $\sim$ 0.003 au \citep{reiners2010}.  Each system is integrated for 20 Myr, which we find to be a sufficient amount of time to fully process the planet forming material into a stable configuration of 4-10 planets.  Out of $\gtrsim$ 800 complete simulations, we hand-select 30 systems that best resemble TRAPPIST-1 in terms of forming exactly seven total planets with orbital period and mass distributions similar to those of the real system. Note that many of the discarded system have 5, 6, 8, or 9 planets formed or have seven total planets but with final masses that were either too high or too low.

Note that the average number of planets formed in our simulations is 5.69 if a ``planet” is defined as having at least the mass of Mercury. The degree of multiplicity in our simulations is most sensitive to the radial extent of the disk and its total mass (see e..g, \citealt{Chambers2001Icar}).  In this sense, our simulations are tuned to form large numbers of planets, however it is still challenging to form 7 planets with sufficiently compact orbits (even with somewhat fine-tuned initial conditions, see also \citealt{coleman2019}).  The degree of multiplicity is only slightly sensitive to the minimum mass used to define planets. Indeed, the average number of planets is 5.33 if the minimum planet mass is defined as 0.3 $M_\oplus$.  For a limit of 0.15 $M_\oplus$, we find an average number of planets of 5.66.  In \citet{Clement2025}, we defined ``Trappist analog systems” as systems that formed between 6-8 planets (225 of 631 total simulations). Of these, only 104 formed exactly 7 planets.

\paragraph{Extended Disks} To place the volatile inventories of the TRAPPIST-1-like systems in the context of different disk radii, disk masses, and stellar masses, we extend our suite of N-body simulations to include stellar masses in $\{0.1, 0.35, 0.7\}~ M_{\odot}$ and disk radii in $\{0.5, 5\}$ AU; we set the total mass of the disk at $\{5, 10\}~M_{\oplus}$ for disks with radii of $0.5$ au, and $\{10, 100\}~M_{\oplus}$ for disks with radii of $5$ au. As in our TRAPPIST-1 simulations, half of the disk mass is uniformly distributed in planet embryos that perturb and interact with all the bodies in the system, and the other half is uniformly distributed in smaller planetesimals that are restricted to only interacting with the embryos. $\{\text{Number of big bodies}, \text{number of small bodies}\}$ is $\{20, 1000\}$ for 5$~M_{\oplus}$ disks, $\{40, 2000\}$ for 10$~M_{\oplus}$ disks, and $\{100, 2000\}$ for 100$~M_{\oplus}$ disks respectively. All bodies in the system are distributed according to a power-law surface density profile $\Sigma(r) \sim r^{-3/2}$, an approximation followed from the ``minimum-mass solar nebula" model \citep{Weidenschilling77}. Initial eccentricities and inclinations of all bodies follow Rayleigh distributions with modes of 0.01 and $1^\circ$ respectively. The remaining angular orbital elements are distributed randomly from 0 to 360 degrees. We simulate each initial configuration until the system has reached stability (typically 10 Myr for 0.5 au disks and 200 Myr for 5 au disks). We simulate the suite of 12 ($3 \text{ stellar masses }\times 2 \text{ disk radii }\times 2 \text{ disk masses }$) different initial configurations 10 times respectively, resulting in a total of 120 different accretion outcomes.

\subsection{Volatile Growth Simulator}

We use accretion histories from a suite of $N$-body simulations to drive a volatile growth simulator (\texttt{VGS}; Figure~\ref{fig1_summary}), adapted from the open-source framework of \citet{Chen+Jacobson22}. To first order, \texttt{VGS} captures the dominant physical processes that govern volatile evolution during the assembly of rocky planets, as described by conventional accretion theory. Although some secondary effects are neglected (see below), previous applications of this model class have successfully reproduced Earth’s present-day inventories of both major and minor atmophile species, supporting the validity of the model and its assumptions.

Each simulation tracks the collisional evolution of a planetary embryo (the parent body) that ultimately forms a protoplanet. \texttt{VGS} post-processes time-resolved outputs from \verb|Mercury6|, including the timing, impact velocities, angles, and heliocentric source regions of all impacting bodies. The model is designed to (i) document the evolving inventories of key volatile elements, hydrogen, carbon, and nitrogen, and their dominant molecular carriers (e.g., H$_2$O, CO$_2$, N$_2$), and (ii) explicitly quantify the transfer of these species across the planetary atmosphere, mantle, and core reservoirs.

By accounting for the timing, composition, and physical characteristics of each collision, \texttt{VGS} enables detailed evaluation of volatile delivery and loss events, including their dependence on stochastic impact histories. This approach allows for the reconstruction of diverse accretion pathways and volatile outcomes relevant for low-mass planets forming in compact systems.

\subsubsection{Accretion of Major Volatile Elements}
\label{sec:method1}

For planetesimal accretion events, we assume that all material is homogeneously mixed into the mantle of the growing parent body. The bulk compositions of both initial embryos and planetesimals are prescribed based on their heliocentric semi-major axis of origin, with initial volatile abundances informed by published cosmochemical reservoir data.

To approximate the compositional gradient across the protoplanetary disk, we divide the disk into three regions, each associated with a distinct meteoritic analog that also corresponds to a dry--wet volatile spectrum:
\begin{itemize}
    \item \textbf{E-type chondrites} (dry, volatile-poor), originating in the innermost disk
    \item \textbf{S-type chondrites} (moderately volatile-bearing), from intermediate distances
    \item \textbf{C-type chondrites} (volatile-rich, water- and carbon-bearing), from the outer disk
\end{itemize}

While tracing chondritic material back to its planetesimal formation location, and forward to the final bulk composition of a formed planet, remains a challenging and uncertain task \citep{alexander2022}, we adopt a simplified but plausible model. Specifically, we implement four step functions to define the bulk composition $\chi_{\rm pl}$ as a function of semi-major axis ($a$), serving as a first-order approximation of heliocentric compositional variation and volatile content:

\begin{equation}
\chi_{\rm pl} =
\begin{cases}
{\rm E{\rm -}type \ (dry,\ volatile\!-\!poor)} & {\rm for} \quad a < a_{\rm in} \\
{\rm S{\rm -}type \ (moderate\ volatiles)} & {\rm for} \quad a_{\rm in} < a < a_{\rm fl} \\
{\rm C{\rm -}type \ (wet,\ volatile\!-\!rich)} & {\rm for} \quad a \geq a_{\rm fl}
\end{cases}
\end{equation}

For our control TRAPPIST-1 simulations, and consistent with the expected location of the snowline around 0.05 AU, we adopt an inner disk boundary of $a_{\rm in} = 0.03$ AU and set the snowline position at $a_{\rm fl} = 0.035$ AU. In ``dry'' disk scenarios, we remove contributions from C-type (carbonaceous) planetesimals altogether. For simulations involving more massive protoplanetary disks ($M_{\rm disk} = 0.35\,M_\odot$), we shift the inner boundary and snowline outward to $a_{\rm in} = 0.15$ AU and $a_{\rm fl} = 0.4$ AU, respectively. For disks that experience an early luminous pre-main-sequence (PMS) phase, characteristic of TRAPPIST-1-like stars, we adjust the snowline to reflect hotter inner disk conditions, with $a_{\rm in} = 0.04$ AU and $a_{\rm fl} = 0.05$ AU. This elevated luminosity effectively moves the snowline outward and limits the contribution of volatile-rich bodies to planets that form inside ${\sim}0.05$ AU. Our N-body simulations using \verb|Mercury6| show that, depending on the surface density profile, many planets accrete exclusively from materials located within this expanded snowline.

We assume nitrogen and carbon to be primarily delivered in their most stable gas-phase forms N$_2$ and CO$_2$, respectively. Table 1 summarizes the adopted abundances of each species across the different chondritic source reservoirs. During each impact event, the total mass and associated volatile inventory of the incoming projectile are incorporated into the growing planet’s atmosphere and mantle reservoirs. Immediately following the accretion, we evaluate atmospheric losses due to impact erosion. Notably, our model does not currently account for volatile partitioning between molten and solid phases of the mantle during post-accretion solidification. For instance, processes like melt entrapment during magma ocean crystallization \citep{carone2025}, where residual melts become sequestered in partially solidified mantle regions, are omitted.

\begin{table}[htbp]
\caption{Initial composition inventory of various cosmochemical reservoirs, expressed as mass fractions. Most volatile carriers in the model exist in condensed phases, primarily organics, rather than as gas-phase species. For instance, carbon is assumed to be present in forms such as graphite, carbonates, and silicon carbide. The model does not simulate direct evaporation of gaseous species during energetic events; instead, volatile loss during protoplanet formation arises solely from impact erosion and energy-limited atmospheric escape. }
   \label{tab:Table 1}
   \begin{center}
   \begin{tabular}{cccc}
\hline
Species & E-type & S-type & C-type\\
\hline
C & $10^{-6}$ & $10^{-4}$ & 0.01 \\
N$_2$ & $5 \times 10^{-7}$ & $10^{-5}$ & $10^{-4}$ \\
H$_2$O & $10^{-5}$ & 0.001 & 0.05 \\
\end{tabular}
\end{center}

\footnotesize{E-type= enstatite chondrite, S-type= ordinary chondrite, C-type=  carbonaceous chondrite. Data taken from \citet{Kerridge1985GCA}, \citet{Robert1982GCA}, \citet{JessbergerEt1988}, \citet{ThomasEt1993GCA}, \citet{AbeEt2000}, and \citet{PearsonEt2006MPS}.}
\end{table}

Melt-trapping mechanisms, such as those occurring during slow crystallization of the mantle, may result in volatile-rich melt pockets that remain isolated from surface exchange for billions of years \citep{noack2017,dasgupta2024}. These processes can cause strong fractionation of refractory elements, depleting the shallow lithospheric mantle while enriching deeper reservoirs in incompatible species. In this sense, our simplified mantle–atmosphere equilibration model is more representative of a shallow convective mantle, and does not track volatiles sequestered in mineral phases such as wadsleyite, ringwoodite, or high-pressure pyroxenes and olivines. Note that these limitations imply that our estimates may slight underrepresent long-term interior volatile retention and thus overestimate the eventual atmospheric and hydrospheric inventories.

\subsubsection{Atmospheric Impact Erosion}
\label{sec:method2}
%observable fingerprints in favor of impacts 
The isotopic composition of Earth’s silicate mantle closely resembles that of chondritic meteorites, yet its elemental abundances are notably depleted in volatiles. This disparity suggests that Earth's atmosphere experienced substantial bulk removal through large impacts rather than through isotopically selective processes such as hydrodynamic escape or mantle ingassing \citep{schlichting+18,sinclair+20}. Indeed, impact-driven atmospheric loss is widely considered a dominant mechanism during the late stages of terrestrial planet formation \citep{WalkerEt1986Icar,Chyba1990NATURE,AhrensEt1993,NiemEt2012Icar}. Collisions during this phase can eject atmospheric gases either through vapor plume expansion \citep{Melosh+Vickery1989} or by transmitting shock waves through the surface that expel gas via ground motion \citep{Genda+Abe2005NATURE}.

To quantify atmospheric loss due to impacts, we implement the formalism of \citet{SchlichtingEt2015Icar}, which treats the collision as an instantaneous point explosion. In this approximation, the kinetic energy of the impactor is rapidly converted into thermal and mechanical energy \citep{melosh1989impact}, generating localized overpressure capable of accelerating and expelling atmospheric mass. The mass of atmosphere lost per impact event, $m_{\rm loss}$, depends sensitively on several parameters including the impactor’s mass, velocity, incidence angle, and the planet’s gravitational binding energy.

For impact velocities comparable to the target’s escape velocity, \citet{SchlichtingEt2015Icar} identified three distinct regimes of atmospheric mass loss, each governed primarily by the size of the projectile relative to the planet’s atmospheric scale height and radius. These regimes enable the scaling of impact erosion efficiency across a wide range of impact conditions, and form the basis of our loss estimates in the volatile growth simulations. These regimes are:

\begin{equation}
m_{\rm loss} =\begin{cases} 
0 &  {\rm for} \hspace{2 mm}   r_{\rm pl} < r_{\rm min} \\
\frac{m_{\rm pl}}{M_{\rm atm}} \left[\frac{r_{\rm min}}{2 r_{\rm pl}}\left[1 - \frac{r_{\rm min}^2}{r_{\rm pl}^2}\right]\right]   &  {\rm for} \hspace{2 mm}   r_{\rm min}  < r_{\rm pl} < r_{\rm cap}  \\
\frac{M_{\rm cap}}{M_{\rm atm}} & {\rm for} \hspace{2 mm}   r_{\rm pl} \geq 2.5
       \end{cases}
\end{equation}

\noindent where $r_{\rm min}$ is the minimum planetesimal radius to eject any mass:

\begin{equation}
    r_{\rm min}=\left(3 \rho_{\rm surf}/\rho_{\rm pl}\right)^{1/3} H_{\rm tar}
\end{equation}

\noindent and $r_{\rm cap}$ is the radius above which all the atmosphere will be lost in the impactor's vicinity:

\begin{equation}
      r_{\rm cap} = \left(3 (2 \pi)^{1/2} \rho_{\rm surf}/4 \rho_{\rm pl}\right)^{1/3} \left(H_{\rm tar} R_{\rm tar}\right)^{1/2}
\end{equation}
 
\noindent $\rho_{\rm surf}$ and $\rho_{\rm pl}$ are the surface density of the planet and the density of the planetesimal respectively, $H_{\rm tar}$ and $R$ are the scale height and the current radius of the parent proto-planet, and $M_{\rm cap}$ is the maximum amount of gas that can be ejected above the tangent plane \citep{Melosh+Vickery1989}. The scale height of the target is $(kT_{\rm a})/(\mu m_H g)$ where $k$ is the Boltzman's constant, $T_{\rm a}$ is the atmospheric temperature which we have assumed to be isothermal, $\mu$ is the time-evolving mean molecular weight, $m_{\rm H}$ is the mass of a hydrogen atom ($1.67\e{-24}$ g), and $g$ is the gravitational acceleration. To determine the evolving radius of the embryo $R$, we use the planetary embryo mass-radius relationship following \citet{JacobsonEt2017EPSL}.

Highly energetic collisions can also generate global-scale seismic ground motions, which in turn produce powerful atmospheric shock waves. These shocks may propagate through substantial portions of the atmosphere, resulting in the ejection of atmospheric mass beyond the planet’s gravitational field. For temperate, Earth-like planets with equilibrium temperatures $T_{\rm eq} \lesssim 400$ K, the atmospheric structure can be reasonably approximated as isothermal. Under this assumption, the atmospheric mass fraction lost during each high-energy impact can be expressed analytically as:

\begin{equation}
\chi_{loss} = 0.4 \left(Y\right) +1.4 \left(Y\right)^2 -0.8 \left(Y\right)^3
\end{equation}

\noindent where $v_{\rm imp}$ is the impact velocity, $v_{\rm esc}$ is the escape velocity, $m$ is the mass of the projectile, $M$ is the mass of the parent proto-planet, and $Y = (v_{\rm imp} m)/(v_{\rm esc} M)$. We apply the above formalism to calculate the mass-lost per impact for large embryos, or for impactors that are greater than 200 km in radius.

%\section{EUV Hydrodynamic Escape}

%					Feuv=29.7*(time/1e9)**(-1.23)*(1.23)**(-2)
%					m_elim = (np.pi*Feuv*target_r**3)/(cgrav*target_m)

\begin{table}[htbp]
\caption{Values of Henry's Law Coefficient and Constants.}
   \label{tab:Table 2}
   \begin{center}
   \begin{tabular}{ccc}
\hline
Species & $K_{\rm H, ref}$ & $C$ \\
  & ($\frac{{\rm mol}}{{\rm kg} \cdot {\rm bar}}$) &   \\
\hline
N$_2$ & $6.1 \times 10^{-4}$ & 1300 \\
CO$_2$ & $3.4 \times 10^{-2}$ & 2400 \\
H$_2$O & $7.8 \times 10^{-4}$ & 500 \\
\end{tabular}
\end{center}

\footnotesize{Data from \citet{adamson1967physical}, measured at standard state temperature of 298 K.}
\end{table}

\subsubsection{Mantle-Atmosphere Exchange}
\label{sec:method3}
Impacts between embryos and planetesimals are often directly associated with magma ocean (MO) formation, which promotes exchange between the atmosphere and underlying mantle.
To estimate the equilibration of each gaseous species between the mantle and the atmosphere after each accretion event, we employ a version of Henry's law given by the following:

\begin{equation}
[A({\rm aq, i})] =  K_{\rm H, i} \left(P_{\rm tot} - h/100 P_{\rm o}\right) f_{\rm i} \label{eq:henry}
\end{equation}

\noindent where $[A({\rm aq})]$ is the concentration of gaseous phase of species i, $K_{\rm H, i}(T)$ is the Henry solubility coefficient, $P_{\rm tot}$ is the total atmospheric pressure, $h$ is the relative humidity, $P_{\rm o}$ is the vapor pressure at ambient $T$, and $f_{\rm i}$ is the mole fraction of gas i in dry air.

Pressure $P = (g M_{\rm atm})/(4 \pi R^2)$ is the surface pressure at the base of the atmosphere, where $M_{\rm atm}$ is the mass of the atmosphere and $R$ is the radius of the proto-planet, calculated from the mass-radius relationship \citep{JacobsonEt2017EPSL}.

$K_{\rm H, i}(T)$, the Henrinian solubility coefficient, is determined by the following equation:

\begin{equation}
K_{\rm H} = K_{\rm H, ref}   \hspace{1 mm} {\rm exp} \left(C \left(\frac{1}{T} - \frac{1}{T_{\rm ref}}\right)\right)
\end{equation}

\noindent where $K_{\rm H, ref}$ is the standard reference solubility constant for species $i$, $C$ is a constant in Kelvins, $T$ is the surface temperature at the base of the atmosphere which we fix at 1500 K (This assumption is valid in that the effects on  bulk composition is not substantially large, which is also noted by \citep{Pepin1997Icar}), and $T_{\rm ref} = 298$ is the standard state temperature. The values for $K_{\rm H, ref}$ and $C$ used are specified in Table~2. 

To find the equilibrium mass to be established in the mantle, we modify Equation~\ref{eq:henry} to the form:
\begin{equation}
m_{\rm gas, eq} =  K_{\rm H, i} P_i  M_i m_{\rm mantle}
\end{equation}

\noindent where we have replaced $\left(P_{\rm tot} - h/100 P_{\rm o}\right) f_i$ with $P_i$. $M_i$ is the molecular mass of species $i$ and $M_{\rm mantle} = 0.6 M_p$ is the current mass of the growing mantle. Whether degassing or ingassing occurs depend on both the amount of gas in the mantle and atmosphere. If $m_{\rm gas, eq}$ is greater than the current amount in the mantle, then we release the difference to the atmosphere and vice versa.

%The majority of Venus's water may have been lost prior to solidification.

\begin{figure*}
\begin{center}
\includegraphics[width=2.\columnwidth]{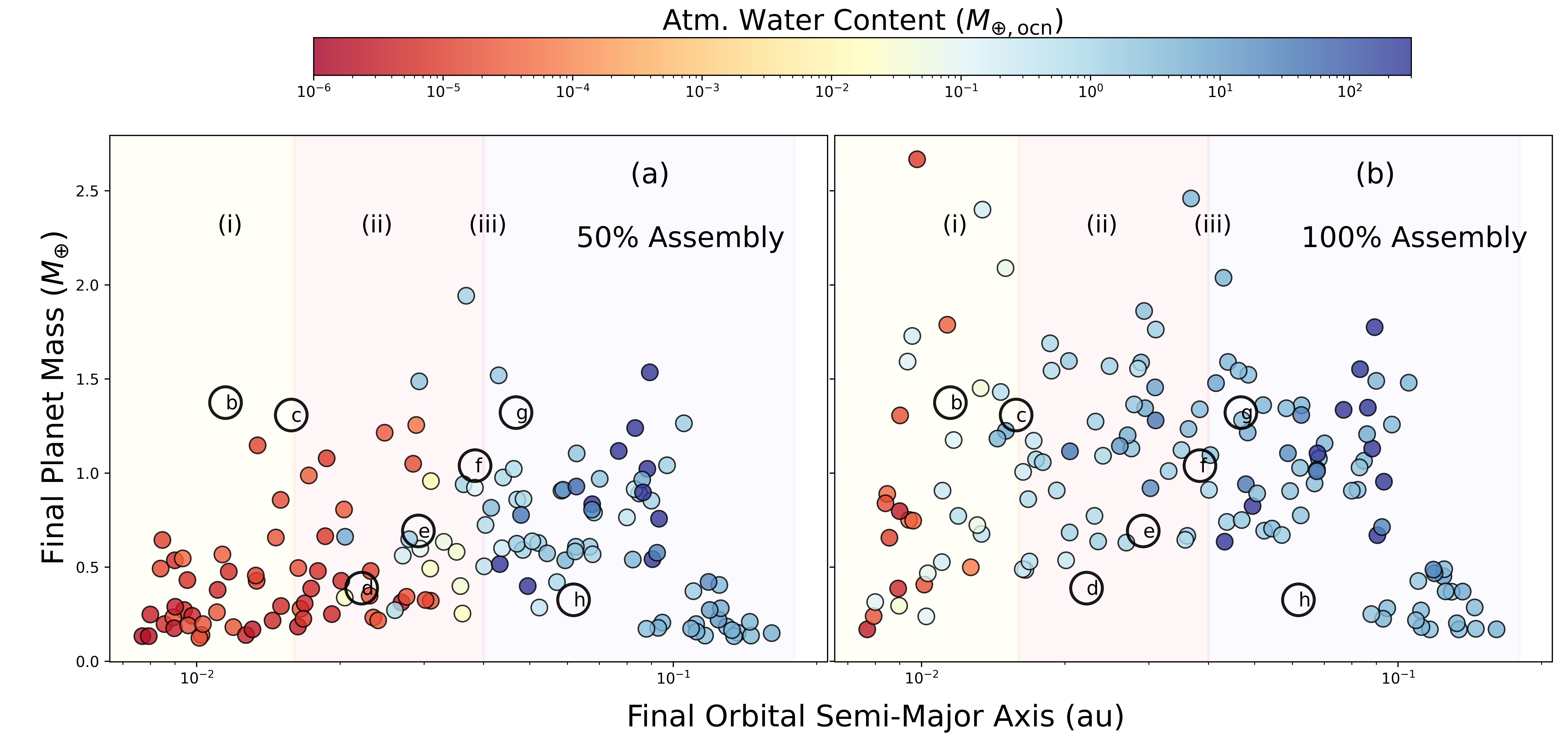}
\caption{\label{fig2_pop} Snapshots of \texttt{VGS}-calculated surface water mass (in units of Earth’s ocean mass) for TRAPPIST-1 analogs, shown as a function of total planet mass ($M_\oplus$) and final orbital distance (AU). Panel (a) depict planets at the halfway point of accretion (i.e., when the protoplanet reaches half its final mass), while panels (b) show the final planet properties. Measured properties of the TRAPPIST-1 planets are indicated by black circles \citep{Agol+21}. At mid-assembly, most inner planets are water-poor due to early impact-driven loss and higher irradiation. By the end of accretion, increased volatile delivery events from more water-rich materials and reduced erosion allow many planets, particularly beyond $\sim$0.03 AU,to accumulate and preserve large water reservoirs, leading to the strong radial gradient in water content.} 
\end{center}
\end{figure*}

\subsubsection{Core Sequestration of Carbon}
\label{sec:method4}
Apart from outgassing to the nascent atmosphere, core formation processes could sequester large portions of siderophiles such as carbon due to its high affinity to metals. To calculate equilibration between the lower mantle and core, we use a simple partition coefficient:

\begin{equation}
D_i = \frac{C^{\rm alloy}_i}{C^{\rm sil}_i}
\end{equation}

\noindent in this equation, $C^{\rm sil}_i = M^{\rm sil}_i / m^{\rm sil}  $ and $C^{\rm alloy}_i = M^{\rm alloy}_i / m^{\rm alloy} $ is the evolving concentration in the mantle and core respectively, where $M^{\rm sil}_i$ is the volatile component $i$ in the silicate, $m^{\rm sil}$ is the mass of the magma ocean, $ M^{\rm alloy}_i$ is the volatile component $i$ in the alloy, and $m^{\rm alloy}$ is the total mass of the core. We follow \citet{deguen2011experiments} to determine the $m$ that interacts with the metal alloy, i.e., from the expression of the melt volume:

\begin{equation}
V_m = \gamma V_{\rm imp} \frac{8\pi G \rho_p R^2_p}{3 E_m}
\end{equation}

\noindent where $\gamma = 0.15$ is a proportionality constant, $V_{\rm imp}$ is the impactor volume,  $G$ is the gravitational constant, $\rho_p$ is the mean density of the proto-planet, $R_p$ is its radius, and $E_m = 9 \times 10^{16}$ g$^2$ m$^2$ s$^{-2}$ is the specific energy. In this study, we do not include the partitioning of nitrogen and hydrogen into metallic phases during core formation. Among volatile elements, carbon has the most robust experimental constraints, and as such, it is the only element we explicitly track during core sequestration. This simplification allows us to manage the complexity of the model while maintaining fidelity in the most well-characterized volatile behavior. Incorporating core partitioning of additional elements such as nitrogen, hydrogen, and oxygen would likely reduce the net volatile flux to the mantle and surface reservoirs, ultimately yielding drier inner planets than those presented in our current results.

\subsection{Hydrodynamic Evaporative Mass-Loss}

To approximate the first-order time evolution of atmospheric escape on close-in exoplanets orbiting chromospherically active M-dwarfs, we adopt the energy-limited escape equation. In this scheme, high-energy stellar irradiation, particularly in the extreme ultraviolet and soft X-ray regimes (XUV; 200–911 $\angstrom$), deposits energy into the upper atmosphere of a planet. This heating drives hydrodynamic expansion of the atmosphere, enabling gas to escape the planet’s gravitational potential well.

The rate of energy-limited mass loss is given by \citep{Watson1981, Murray-ClayEt2009ApJ, erkaev2016}:

\begin{equation} 
\frac{{\rm d} M_p}{{\rm d} t} = - \frac{\epsilon_{\rm euv} \pi F_{\rm euv}  R_{p}^3}{G M_p}
\end{equation}

\noindent where $\epsilon_{\rm EUV}$ is the mass loss efficiency, i.e., the fraction of incident EUV energy that contributes to unbinding atmospheric gas, which depends on both the atmospheric composition and the stellar flux. We adopt a fiducial value of $\epsilon_{\rm EUV} = 0.1$, consistent with prior modeling of rocky exoplanets and proto-atmospheres \citep{Jackson2012}. $M_p$ and $R_p$ are the evolving mass and radius of the growing planetary embryo, where $R_p$ is determined using the embryo mass-radius relationship from \citet{JacobsonEt2017EPSL}. $G$ is the gravitational constant.

The stellar EUV flux, $F_{\rm EUV}$, is modeled as a power-law function of time and distance, following empirical calibrations for M-dwarfs \citep{RibasEt2005ApJ, ValenciaEt2010A&A}:
\begin{equation} 
F_{\rm euv} = \alpha t^{-\beta}a^{-2}
\end{equation}

\noindent where $t$ is the stellar age in gigayears, $a$ is the planet’s semi-major axis in AU, and $\alpha$ and $\beta$ are constants that characterize the host star’s spectral type and EUV luminosity evolution. For TRAPPIST-1, a well-studied ultracool dwarf, we adopt $\alpha = 67.2~\rm erg~s^{-1}~cm^{-2}$ and $\beta = 0.87$, values broadly consistent with empirical estimates of M-dwarf EUV activity over gigayear timescales.

\begin{figure*}
\centering
\includegraphics[width=2.\columnwidth]{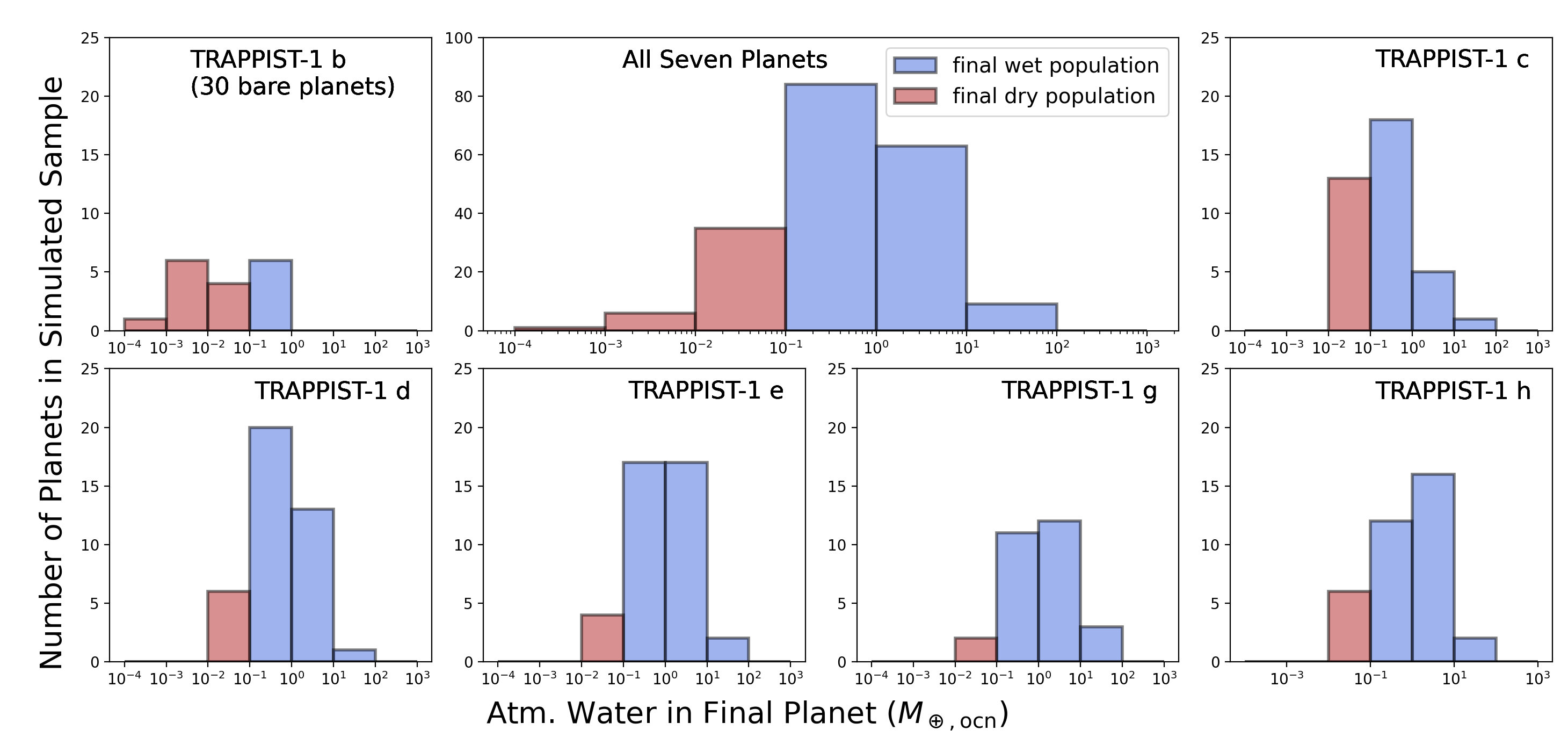}
\caption{\label{fig3_hist} Histogram of simulated TRAPPIST-1 analogs, binned by final atmospheric water content (in Earth ocean masses, $M_{\oplus,\mathrm{ocn}}$). Blue bars denote water-rich worlds capable of sustaining at least a shallow global ocean ($>10\%$ of Earth’s surface water), while red bars represent water-poor worlds ($<10\%$). Each panel corresponds to a specific TRAPPIST-1 planet, with the center panel showing the combined distribution for all seven.
Across the ensemble, inner planets (e.g., b–c) are predominantly dry, whereas mid-to-outer planets (e.g., e–h) are skewed toward water-rich outcomes, reflecting the strong radial dependence of volatile retention identified in our simulations.}
\end{figure*}

\section{Results}

To investigate the conditions under which rocky exoplanets in compact systems like TRAPPIST-1 acquire and retain volatiles, we carry out a suite of coupled $N$-body and volatile-growth simulations that span a range of disk compositions, stellar luminosity histories, and system architectures. By varying disk surface density profiles and the chemical makeup of accreting material, we explore plausible evolutionary pathways for volatile delivery and loss. Our model explicitly tracks the histories of water, carbon, and nitrogen during planetary accretion and evaluates their final incorporation into the bulk inventories of TRAPPIST-1 analogs. In the sections that follow, we compare outcomes across different planetary growth timescales, both with and without the influence of a luminous pre-main-sequence (PMS) phase, and assess how planetary mass, orbital location, and system configuration jointly govern volatile retention.

\subsection{Evolution and Final Water Inventories on Nascent TRAPPIST-1 Worlds}
\label{results1}

Our \texttt{VGS} simulations indicate that more than 70\% of the inner TRAPPIST-1 analog planets (b and c) form  dry, with the majority retaining less than $0.1M_{\oplus,\rm ocn}$. In contrast, the outer planets (g and h) exceed this threshold in over 80\% of cases, giving them a high likelihood of attaining Earth-like water mass fractions. Figure\ref{fig2_pop} summarizes these outcomes across all seven planets at two representative epochs, showing their masses, orbital semi-major axes, and surface water inventories (expressed in $M_{\oplus,\rm ocn}$, where 1 $M_{\oplus,\rm ocn}$ is the mass of Earth’s present-day oceans).

The three-color shading in the same figure delineates the three potential climate regimes for the seven TRAPPIST-1 planets: runaway (i), temperate (ii), and cold (iii), corresponding roughly to orbital separations $<0.015$ au, $0.015$–$0.04$ au, and $>0.04$ au, respectively. Given the inherent stochasticity of N-body accretion models, it is important to note that some simulations yield planetary architectures quite dissimilar to that of the actual TRAPPIST-1 system. In fact, the most statistically “typical” systems formed under our initial conditions do not resemble TRAPPIST-1. If we require $M_p \ge 0.3~M_\oplus$, the mean number of planets per system is 5.33; only 225 out of 631 simulated systems formed 6–8 planets, and just 104 yielded exactly seven planets. As such, we hand-select planet analogs that most closely resemble the real system, though certain features (such as precise orbital spacing) remain difficult to reproduce (see \citealt{Clement2025} for details). Across both dry and wet initial conditions, a planet’s current orbital semi-major axis is a strong predictor of its total water content, as evidenced by the clear red-to-blue gradient in Fig.~\ref{fig2_pop}.

Simulations corresponding to the regimes of planet analogs b and c (region i) exhibit the widest range of planetary outcomes, both in final planet mass and surface water abundance. Even with water-rich planetesimals, fewer than 20\% of inner planets retain surface water levels exceeding 0.1 $M_{\oplus,ocn}$ (10\% of Earth’s ocean mass). In contrast, temperate-zone cases in region ii (planets d, e, and f) yield the largest number of modeled planets with Earth-similar ocean inventories, in these zones, 60–80\% of cases have 0.5–5 $M_{\oplus,ocn}$. For the cold outer planets g and h (region iii), only a few simulated cases (<10\%) result in more than $10\times$ Earth’s ocean mass. Notably, smaller planets with masses below $\sim0.5~M_\oplus$ are systematically drier than their larger and more distant counterparts. When formed from dry planetesimals, planets in this mass range most often accrete between 0.01 and 0.1 $M_{\oplus,ocn}$ (not shown).

In terms of mass dependence, our simulations indicate that the most water-abundant planets tend to have intermediate masses, while both the smallest ($M_p < 0.6~M_\oplus$) and the largest planets ($M_p > 2.0M_\oplus$) are more often drier. The lowest-mass planets (b and h analogs) are typically dry due to insufficient impact-driven delivery. Conversely, most planets above $2.0M_\oplus$ lose a substantial fraction of their accreted volatiles through giant impacts and photoevaporation, with the escape rate scaling as $R_p^3$, although a few high-mass cases retain $>1~M_{\oplus,ocn}$. Due to the stochasticity of accretion histories and radial variations in the water mass fraction (WMF) of the protoplanetary disk, we constrain the likelihood that TRAPPIST-1 planets possess Earth-like volatile contents by analyzing a large ensemble of time-resolved simulations. Figure~\ref{fig3_hist} summarizes the surface water distributions across both wet and dry initial conditions. With the exception of planets in the innermost regions (b and c analogs), planets in the temperate and cold regimes have $\gtrsim$50\% probability of achieving Earth-like ocean masses.

Overall, the majority of TRAPPIST-1 analog planets exhibit a non-negligible probability ($\geq$20\%) of acquiring near Earth-like ocean masses. This outcome arises from the ability of even the innermost planets to accrete material originating beyond the snowline under certain disk surface density profiles. In particular, steeper profiles facilitate the inward scattering of volatile-rich planetesimals onto the orbits of b and c analogs, while simultaneously reducing the atmospheric erosion caused by smaller colliders (see \citealt{Clement2025}). Interestingly, these same disk profiles also tend to reproduce other observed system characteristics, such as the overall mass distribution and orbital spacings. Although delivery efficiency is lower for these inner planets, the shorter radial distance for planetesimal migration increases the probability of interception.

By contrast, TRAPPIST-1 h analogs generally emerge as smaller, drier planetary embryos stranded beyond the more massive planets f and g, consistent with the real system and reminiscent of Mars’s inferred formation pathway \citep{dauphas2011}. Assuming TRAPPIST-1-like systems exhibit a range of initial disk WMF gradients, our results, mixing both dry and wet planetesimal populations, serve as a statistical guide to the expected volatile contents of planets in similar compact, low-mass systems.

\begin{figure*}
\centering
\includegraphics[width=2.\columnwidth]{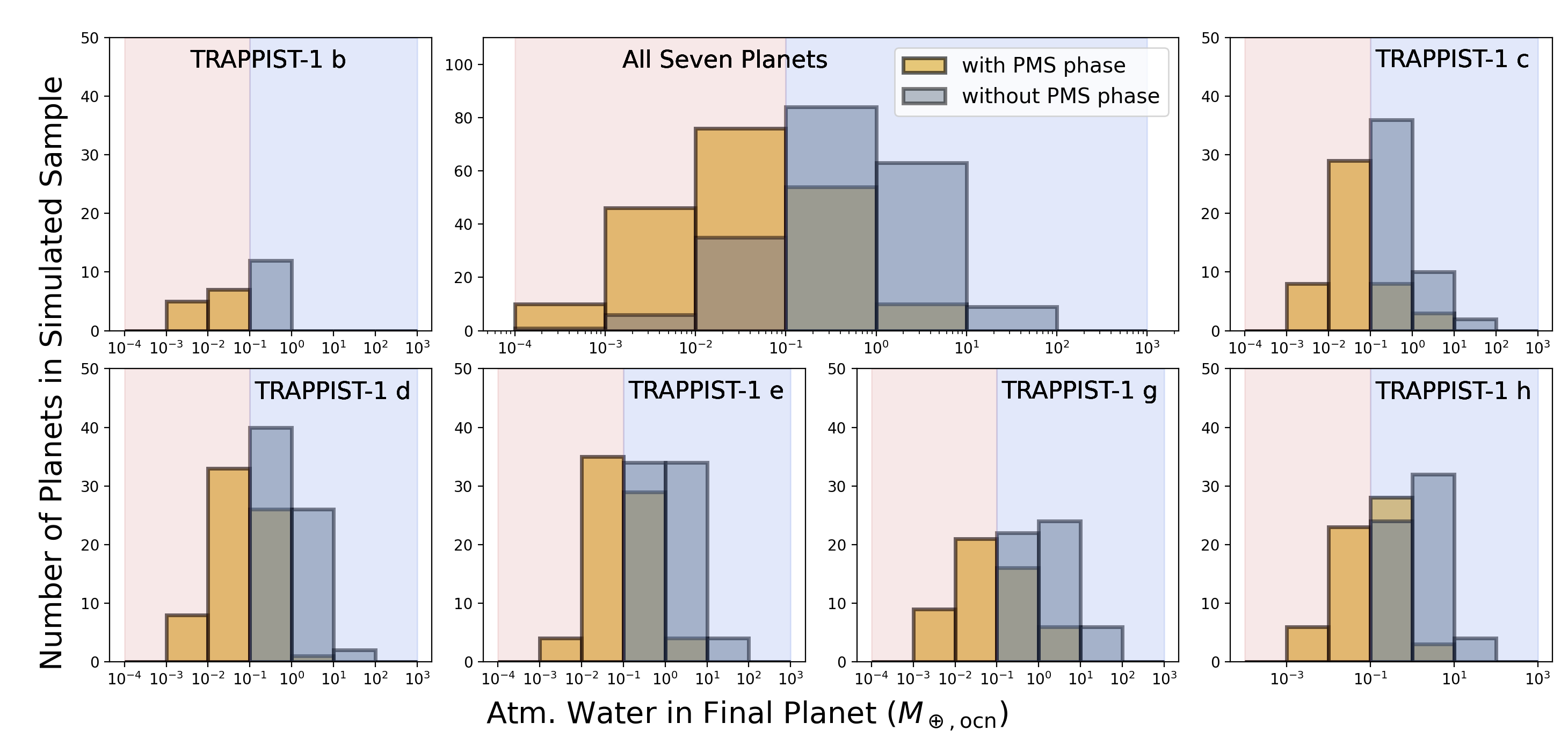}
\caption{\label{fig4_pmshist} Histogram for the occurrence of final planets across two distinct populations of TRAPPIST-1 analogs, binned by their surface water mass (in units of Earth ocean mass). Colors denote whether the initial materials are assumed to have formed in a system with the inclusion of a luminous pre-main sequence (PMS) phase (gold) or without (gray). Color shading separates planets capable of sustaining at least a shallow global ocean ($10\%$ of Earth’s surface water mass) from those considered ``dry.” Across most planets, simulations without a PMS phase yield wetter outcomes, particularly for inner planets, highlighting the strong role of early stellar luminosity in depleting primordial volatiles.}
\end{figure*}

\subsection{Inclusion of a Bright Pre-Main-Sequence Phase}
\label{results4}

The position of the snowline plays a dual role: it not only sets the initial volatile distribution within the protoplanetary disk but also regulates the exposure of young planets to the extreme ultraviolet (EUV) flux that drives hydrodynamic escape. For M dwarfs, the extended pre-main-sequence (PMS) phase, lasting up to several hundred Myr \citep{baraffe2015}, can push the snowline much farther outward during the earliest stages of planet formation.

To test this effect, we ran an additional suite of models using a distinct set of initial preplanetary compositions while preserving the same N-body accretion outcomes. Including a luminous PMS phase, effectively shifting the snowline outward, produces marked changes in the final surface water content across all seven planets compared to the control population in Section~\ref{results1}. As shown in Figure~\ref{fig4_pmshist}, enhanced early disk heating dehydrates solids in the inner regions, yielding planets that are on average 1–2 orders of magnitude drier (gold histograms) than those formed under a snowline location appropriate for a main-sequence ultracool dwarf (gray histograms). This shift is most pronounced for planets b–e, where PMS runs are almost entirely confined to the low-water bins, and is smaller but still visible for the outer planets g and h. The most striking change is the sharp drop in the number of planets retaining more than $10^{-1}~M_{\oplus, \rm ocn}$ of surface water, threshold levels for sustaining shallow global oceans. The All Seven Planets panel in Figure~\ref{fig4_pmshist} illustrates this drop in aggregate, with PMS runs showing far fewer planets above this threshold. This suggests that planetary systems forming during an extended PMS phase are predisposed toward intrinsically dry rocky planets.\footnote{These calculations use extreme end-member scenarios to bracket the plausible range of volatile contents. In reality, the volatile inventory of a TRAPPIST-1-like system would depend sensitively on the host star’s detailed luminosity history.}

\begin{figure*}
\centering
\includegraphics[width=2.\columnwidth]{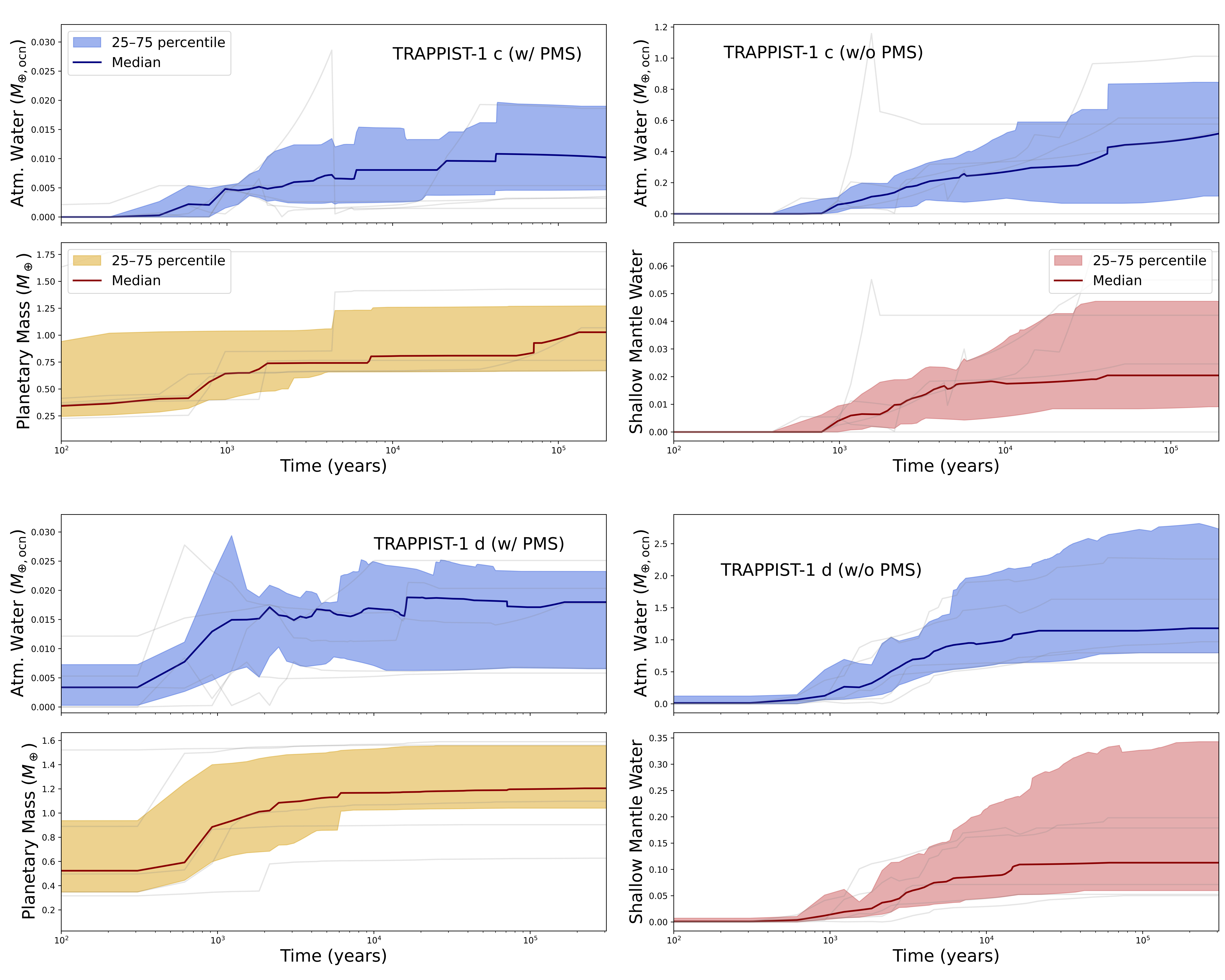}
\caption{\label{fig5_massts} Time evolution of \texttt{VGS}-calculated atmospheric water, shallow mantle water, and planetary mass growth for TRAPPIST-1 c and TRAPPIST-1 d analogs. Solid lines show the median values across simulations, with shaded regions marking the 25th–75th percentile range. Light, thin curves represent five randomly selected individual runs. In both planets, the inclusion of a luminous pre-main sequence (PMS) phase (left panels) strongly suppresses atmospheric water growth compared to PMS-free cases (right panels), with final water contents differing by up to a factor of three. This suppression is most pronounced in the inner planet (TRAPPIST-1 c), where PMS heating limits early volatile retention. Shallow mantle water is shown only for PMS-free scenarios, as PMS-driven volatile loss prevents substantial mantle hydration in the PMS-included runs.}
\end{figure*}

The volatile content of a planet is influenced not only by its initial bulk composition but also by the coupled effects of stellar irradiation history and the timing of accretion. These dependencies are particularly evident when comparing the time-resolved H$2$O inventories of neighboring planets TRAPPIST-1 c and d (Figure~\ref{fig5_massts}). In simulations without the inclusion of a pre-main sequence (PMS) phase, both planets exhibit steep and sustained increases in atmospheric and shallow mantle water, with the bulk of accretion occurring between $10^4$ and $10^5$ years. This rapid growth is seen in the widening of the shaded percentile bands and the steep slope of the median curves in the no-PMS panels. In this scenario, median TRAPPIST-1 c analogs accumulate up to ${\sim}0.5~M{\oplus, \rm ocn}$ of atmospheric water, while TRAPPIST-1 d analogs approach Earth-like levels, with maxima exceeding $2.5~M_{\oplus, \rm ocn}$.

By contrast, including a luminous PMS phase suppresses volatile delivery. PMS heating reduces the inward supply of volatile-rich solids, flattening accretion curves and lowering final water inventories. TRAPPIST-1 c analogs under PMS conditions end up with median atmospheric water masses an order of magnitude smaller than in the non-PMS case. Likewise, shallow-mantle reservoirs in both c and d grow more slowly and to substantially smaller values when PMS heating is included (lower-right panels, Figure~\ref{fig5_massts}). Here, the suppression is visible not only as a lower median but also as a narrowing of the shaded percentile ranges, reflecting a reduced spread in volatile outcomes. In both PMS and non-PMS cases, early, rapid accretion events are rare outliers; most systems follow smoother, delayed volatile buildup. Together, Figures~\ref{fig4_pmshist} and \ref{fig5_massts} show that PMS heating both reduces the final volatile inventory and alters the temporal pathway by which volatiles are acquired. These results indicate that the PMS phase fundamentally reshapes the thermal and chemical landscape of compact planetary systems, limiting the long-term retention of volatiles such as water.

\section{Discussion \& Conclusion}
\label{discussion}

Our model incorporates the key processes that govern planetary accretion and proto-atmospheric formation. Across simulations both with and without the luminous pre-main-sequence (PMS) phase, we find that the hottest rocky planets acquire only minimal volatile inventories. This early-stage depletion offers a natural pathway to intrinsically airless worlds, potentially explaining the growing number of exoplanets without detectable atmospheres in JWST thermal emission observations \citep{lustig2023,greene2023,zieba2023}: planets assembled under such thermal regimes may be inherently dry and unable to sustain surface H$_2$O (e.g., \citealt{miyazaki2022}). Even in scenarios with strong radial mixing, where volatile-rich planetesimals from beyond the snowline are injected inward, volatile delivery to the innermost planets remains inefficient. While photoevaporation has been widely invoked to explain the loss of close-in super-Earth and sub-Neptune atmospheres \citep{owen2013}, our results point to a complementary pathway: many rocky planets may emerge from the accretion phase already desiccated, placing fundamental limits on the atmospheric retention channels available to them.

This perspective reshapes how we think about post-formation escape processes. While previous studies have shown that XUV-driven escape can erode both primordial and secondary atmospheres from Earth-like planets around M dwarfs \citep{johnstone2019,Kite+Barnett2020PNAS}, such outcomes remain highly sensitive to model assumptions. Our results offer a different angle: if the inner TRAPPIST-1 analogs begin intrinsically dry, then even without atmospheric escape, mantle cycling alone would be unable to resupply sufficient volatiles to build a secondary atmosphere, potentially preventing one from forming in the first place.

Although a small number of stochastic outliers experience rapid water delivery within the first $10^{3}$ years, most planets in our simulations acquire the bulk of their H$2$O between $10^4$ and $10^5$ years. This interval matches theoretical timescales for assembling protoplanets from planetesimals and embryos \citep{kokubo2000}, but in compact systems like TRAPPIST-1, the process is inherently faster than in the Solar System. For example, the innermost TRAPPIST-1 planets orbit at 0.02 AU with periods of only ${\sim}1$–2 days, compared to ${\sim}1$ year at 1 AU, implying growth rates up to $\sim$365 times faster (scaling with the local Kepler frequency, $\Omega \propto P^{-1}$). Oligarchic growth timescales also scale as $\Sigma{\rm solids}^{-1}$ (surface density) and $M_*^{-1/2}$. By shifting a minimum-mass solar nebula inward, $\Sigma$ increases by ${\sim}a^{-3/2}$, further accelerating accretion. From these first-principles scalings, the rapid delivery of volatiles,often within ${\sim}0.1$ Myr, is an expected outcome.

Despite the predicted dryness of the innermost planets (e.g., TRAPPIST-1 b and c analogs), many outer planets in our simulations (e.g., e, f, and h analogs) end with volatile budgets comparable to Earth and other Solar System terrestrials (Figure~\ref{fig2_pop}). This is somewhat counterintuitive, given their accelerated accretion histories and compact orbital spacing. One might expect that the absence of giant planets would suppress radial mixing, yet our models show that steeper disk surface density profiles can enhance the inward transport of volatile-rich material. Notably, these same profiles reproduce multiple observed properties of TRAPPIST-1, mass distribution, planet spacing, and low debris levels, without overproducing outer debris disks, consistent with observational upper limits \citep{marino2020}.

Our model identifies impact-driven delivery and erosion as the primary processes shaping planetary volatile budgets. This approach contrasts with \citet{krissansen2024}, who examined the interaction between primordial nebular gas and planetary interiors under the assumption of substantial primary atmospheres. By comparison, our simulations start from the premise that TRAPPIST-1-like planets either never accreted significant primary envelopes or lost them early. Despite these differing assumptions and modeling frameworks, both studies arrive (albeit by different pathways) at a similar conclusion: inner planets are prone to volatile loss through processes tied to accretion or its aftermath. Our results additionally show that mantle volatile depletion limits the potential for volcanic replenishment, especially for TRAPPIST-1b. Meanwhile, TRAPPIST-1e analogs may retain water inventories consistent with scenarios modeled in previous work (e.g., 10–1000 bars H$_2$O). In the future, integrating chondritic delivery with nebular equilibration models could illuminate the formation of second-generation planets in post-main-sequence environments (e.g., \citealt{zhan2024,shields2025,baker2025}).

Our results show that neglecting impact erosion can substantially overestimate final volatile inventories, particularly for outer planets. Disk evolution and population synthesis studies \citep{kimura2022, muller2024, childs2023} often predict a broad range of water mass fractions, with some outer planets retaining 10–20\% of their mass in water. However, these models typically omit or oversimplify the role of impacts in volatile loss. By explicitly incorporating a spectrum of impactor sizes, from kilometer-scale bodies to planetary embryos, our simulations produce more tempered volatile abundances, while still permitting outcomes with tens of percent water by surface mass. The considerable scatter in these results underscores the intrinsically stochastic nature of volatile delivery and erosion.

We do not explicitly model the formation of surface oceans, which would constitute a third volatile reservoir in addition to the mantle and atmosphere. Nonetheless, ocean condensation during accretion could play a significant role. For example, \citet{lock2024} proposed that preferential loss of atmosphere over ocean may establish a positive feedback loop that accelerates atmospheric erosion. In the case of rapidly accreting TRAPPIST-1 planets, however, this effect is likely minimal, as the short intervals between successive impacts would limit the time available for stable ocean formation.

Hydrodynamic escape represents another key source of uncertainty in our models. The energy-limited prescription adopted here can misestimate escape rates by orders of magnitude \citep{krenn+21}, potentially biasing volatile retention predictions. Addressing this will require more sophisticated hydrodynamic or magnetohydrodynamic treatments, particularly to assess how stellar flares influence stratospheric moisture content and drive upper-atmosphere photodissociation \citep{Chen+2021,Chen2025,dong+17}.

We also neglect mantle–atmosphere chemical equilibrium and the presence of reduced species such as CH$_4$, NH$_3$, and CO. This simplification is partly justified by assuming rapid mantle oxidation and minimal hydrogen accretion in these low-mass systems. Moreover, recent work \citep{gu2024} suggests that volatile partitioning differs little between reduced and oxidized interiors, at least for first-order estimates of H$_2$O content. Nonetheless, a full treatment of magma ocean solidification, including its timing \citep{carone2025} and co-evolving outgassing rates \citep{thomas2025}, would enhance the predictive capability of \texttt{VGS} and refine our volatile inventory estimates.

Our model further simplifies core formation by including only carbon sequestration, omitting N, H, and O partitioning into metallic cores. It also does not resolve magma ocean evolution as a function of impact energy. Large impacts, in particular, may drive high-pressure equilibration and alter deep volatile storage. Incorporating scaling laws for melt production and distribution \citep[e.g.,][]{nakajima2021} would allow future studies to capture these processes and assess their influence on the long-term volatile budgets of compact M-dwarf planets.

Finally, our N-body simulations assume in-situ formation, without invoking large-scale migration or pebble accretion. While this assumption is consistent with many of the dynamical and structural properties observed in the TRAPPIST-1 system \citep{coleman2019,childs2023,Clement2025}, it does not exclude the possibility of more distant formation followed by inward migration. Incorporating such alternative formation pathways in future work could test the robustness of our volatile delivery trends and further constrain the dynamical history of compact M-dwarf systems.

\subsection{Data \& Code Availability}
The data supporting the plots and findings of this study are available from the corresponding author upon request. The \texttt{Mercury6} N-body integrator is publicly available at https://github.com/smirik/mercury. A version of the volatile growth simulator (\texttt{VGS}) package can be accessed at https://github.com/skepticalblver/vgs. The most up-to-date version of the code(s) used in this study will be made available upon request.

\software{Mercury6 \citep{chambers1999}; Matplotlib \citep{Hunter2007}; Volatile Growth Simulator (VGS) }

\acknowledgements
This material is based upon work conducted as part of the CHAMPs (Consortium on Habitability and Atmospheres of M-dwarf Planets) team, supported by the National Aeronautics and Space Administration (NASA) under Grant Nos. 80NSSC21K0905 and 80NSSC23K1399, issued through the Interdisciplinary Consortia for Astrobiology Research (ICAR) program. M.S.C. is supported by NASA Emerging Worlds grant 80NSSC23K0868 and by NASA’s CHAMPs team. The N-body simulations presented in this study were supported by the Carnegie Institution and conducted at the Resnick High Performance Computing Center, a facility supported by the Resnick Sustainability Institute at the California Institute of Technology.

H.C. acknowledges the College of Engineering and Science at Florida Institute of Technology for summer research support. H.C. also acknowledges the AI.panther computational facility at Florida Tech, supported by the National Science Foundation MRI Award No. 2016818 (project title: ``Acquisition of a High Performance GPU/CPU Cluster for Research and Innovation in Computational Sciences and Engineering”).

\newpage

\appendix

\subsection{Carbon, Nitrogen, and Accretion Histories}
\label{results2}

Similarly to water contents, our modeling technique is able to provide a predicted suite of carbon and nitrogen accretion outcomes on our nascent proto-planet models,  given  assumptions regarding the range of the compositions of the initial preplanetary materials. As with the simulated final water contents, we find a spread in the CO$_2$ abundance for  each TRAPPIST-1 planet due to disparate initial volatile contents and subsequent accretion histories (or the slightly different slope of the disk’s surface density profile; see Sect.~\ref{sec:methods}), with the largest and smallest planets having accreted the least CO$_2$. Planets that are just slightly more massive than that of Earth have the highest prospects of accreting several times more carbon in the form of CO$_2$. than contained in the Earth. Much like the real system, in our simulations planet b tends to be the most massive final planet, while planet h is almost always the smallest. As such, the former is often found with a carbon-depleted outcome and the latter having more carbon-enriched ones.

\renewcommand{\figurename}{Appendix Figure}
\setcounter{figure}{0}
\begin{figure*}[b] %different options for where to place figure
\begin{center}
\includegraphics[width=1.\columnwidth]{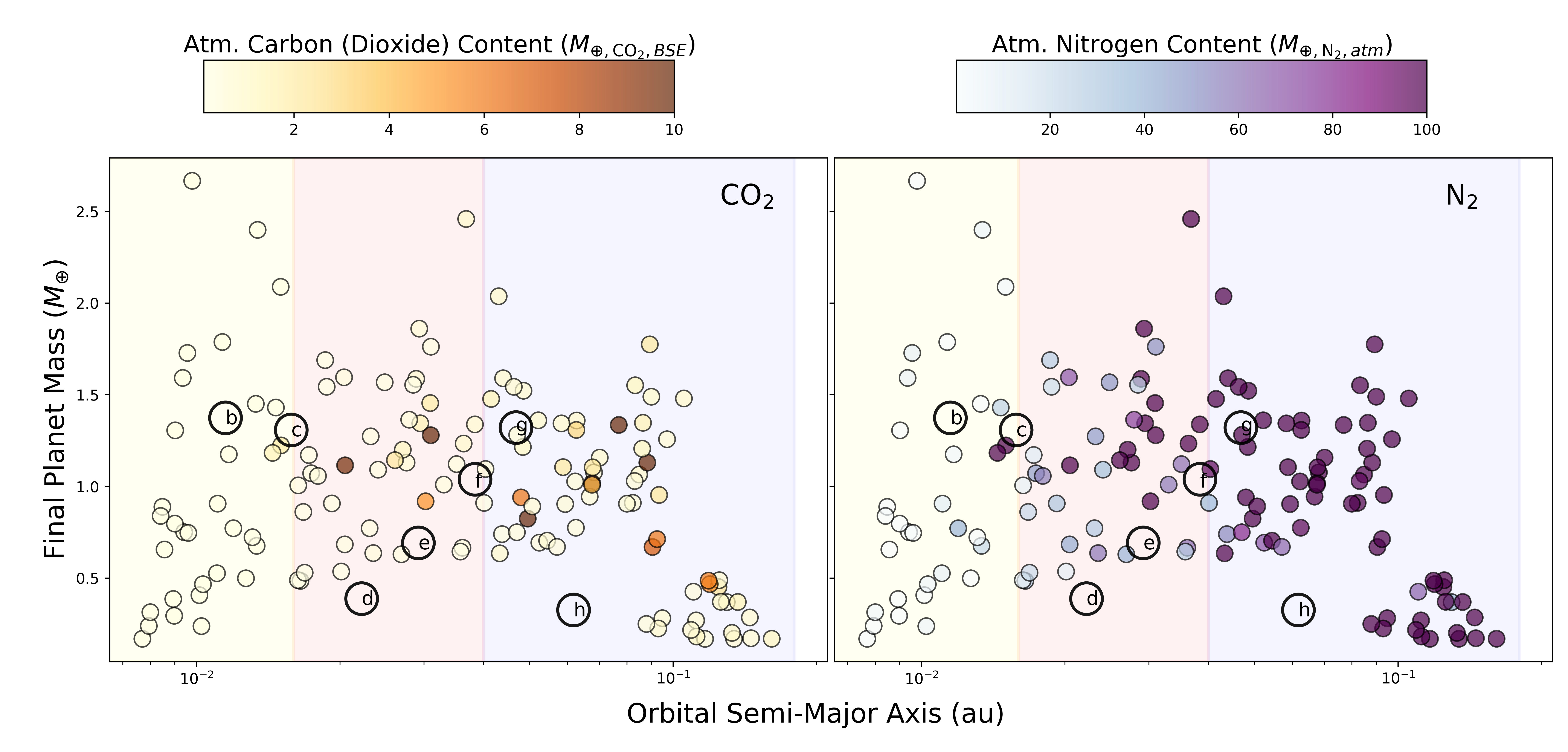}
\caption{\label{fig4_co2n2}   Final atmospheric CO$_2$ (left) and N$_2$ (right) masses of each planet as a function of total planet mass (in units of $M_\oplus$) and orbital distance (in AU). Nitrogen content is expressed in units of $M_{\oplus,N_2,atm}$ , or the mass of N$_2$ in Earth's atmosphere. Carbon content is expressed in units of $M_{\oplus,CO_2,{\rm BSE}}$ , or the total mass of carbon in the BSE (Bulk Silicate Earth). The water mass fractions of accreted planetesimal and embryos assume to follow the step wise function given by Equations~1-3.  Measured TRAPPIST-1 properties are indicated by black circles \citep{Agol+21}. }
\end{center}
\end{figure*}  

Unlike the distributions of water and carbon contents, we find that nitrogen content is a stronger function of orbital semi-major axis and weaker function of planet mass. Only the innermost planets, e.g., planets d and planet e, show nitrogen abundances that are only a few factors greater than that of the bulk Earth nitrogen budget, whereas the colder planets (planets g and h in particular) possess substantially higher nitrogen contents. The nitrogen contents on the inner planets have likely been lost through impact erosion and photoevaporation. This would occur at early epochs during which their nitrogen-rich  atmospheres are lost, leading to a more carbon-enriched atmosphere compared to nitrogen (e.g., \citealt{tucker+14}). Note however, that the Earth is  depleted in N relative to C and H and the cause of this depletion is not understood. Therefore, our predictions of N here may need to be taken with a grain of salt. 

\begin{figure*}[b] %different options for where to place figure
\centering
\includegraphics[width=0.48\columnwidth]{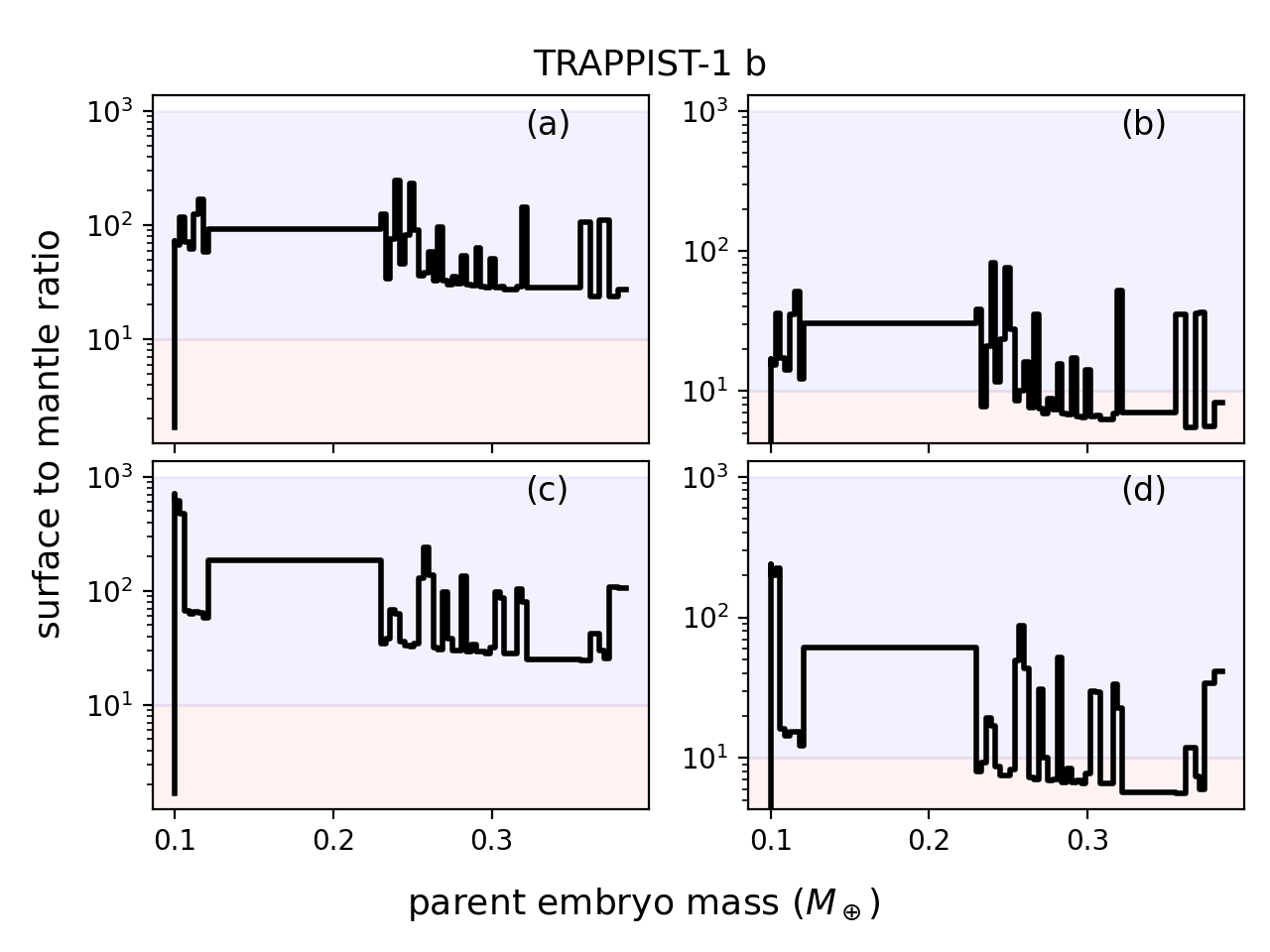}
\includegraphics[width=0.48\columnwidth]{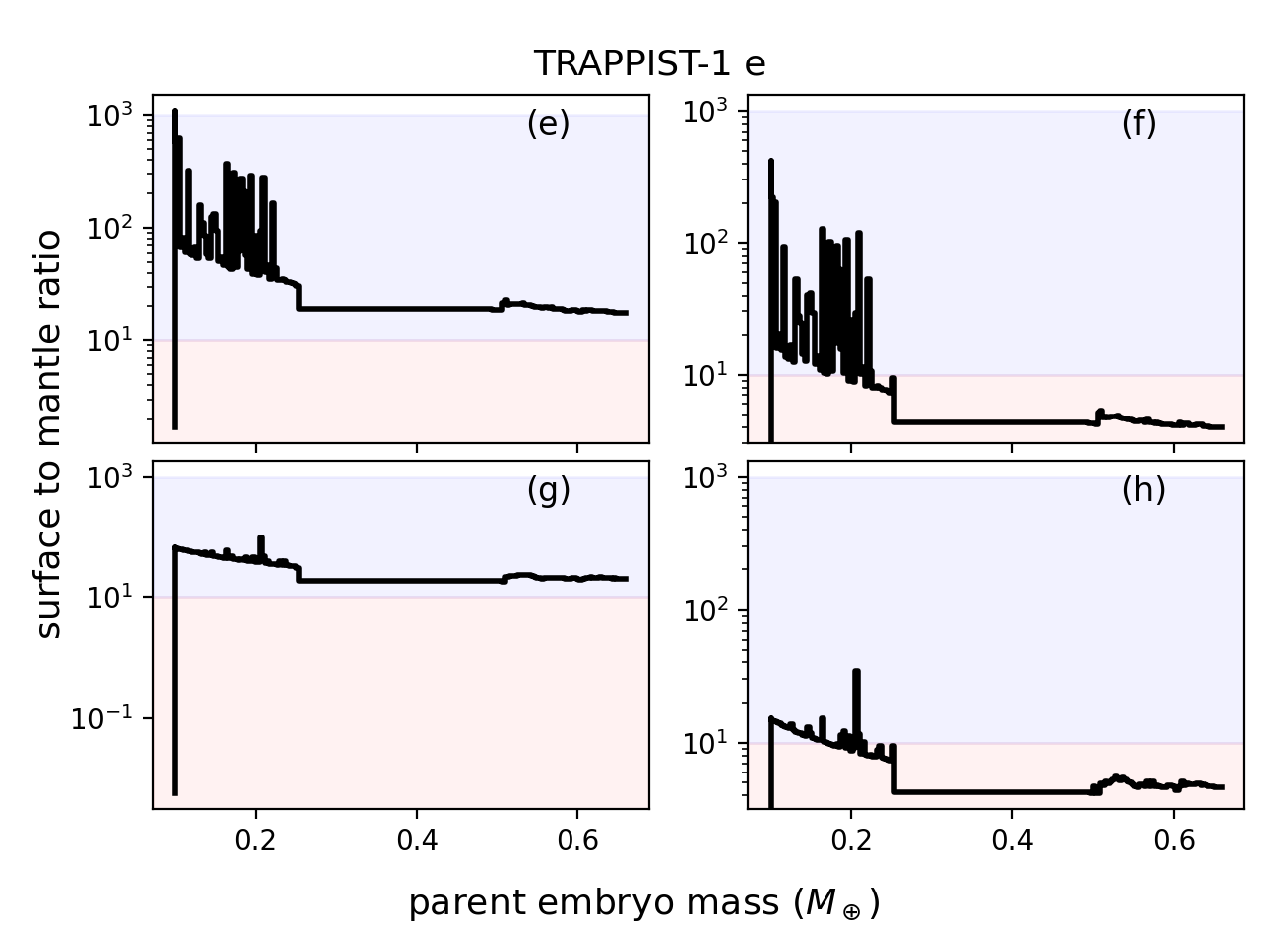}
\caption{\label{fig5_mantatm} Timeseries of \texttt{VGS}-simulated  surface-to-mantle volatile mass ratios for example simulations of TRAPPIST-1 b (a, b, c, d) and TRAPPIST-1 e (e, f, g, h). The panels are: H$_2$O wet planetesimal (a and e), CO$_2$ wet planetesimal (b and f), H$_2$O dry planetesimals (c and g), and CO$_2$ dry planetesimal (d and h). Refer to Sect.~\ref{sec:methods} regarding how the water mass fractions are set in our model. }
\end{figure*}

The ratio of the atmosphere-to-mantle gaseous reservoir provides information on the rates of impact ablation, mantle-atmosphere exchanges, and ultimately, the transfer of the gaseous species of interest. Figure~\ref{fig5_mantatm} shows temporal evolution tracks of the protoplanet's atmosphere to mantle ratio for H$_2$O and CO$_2$  abundances. Our results show that the ratio of the two volatiles are highest at early times and particularly pronounced for planet e. For planet b, the ratios remain relatively constant with large increases in the initial volatile contents of the planet-forming material only producing slight decreases in the atmosphere-to-mantle gaseous reservoir. We find that both species tend to be more dominant in the atmosphere, a trend that is in agreement with our evolving oxygen fugacity and redox state calculations. Though for many cases of our planet e analogs, the ratio approaches unity at the final stages of assembly (Figure~\ref{fig5_mantatm}e-h). While models incorporating multiple gas-phase species and redox conditions can be helpful (see also \citealt{gu2024}), our results place first-order constraints on a subset of traditional volatile species most often invoked when discussing planetary habitability. 

\subsection{Dependence on Stellar Mass and Initial Disk Radius}
\label{results3}

Thus far, our analysis has focused on a fiducial suite of planet accretion histories modeled after the TRAPPIST-1 system. However, compact multiplanet systems span a wide range of orbital architectures and total planetary masses \citep[e.g.,][]{chachan2024}. To place our results in a broader context, we extend our study to explore how variations in protoplanetary disk radius and stellar mass influence planetary surface water content.

In our TRAPPIST-1–like baseline models, volatile accretion strongly depends on a planet’s semi-major axis. When disk properties are varied (Table 3), this dependence becomes less pronounced: total water mass becomes comparably sensitive to both disk mass and radius, with disk mass generally exerting the stronger influence.

For disks truncated at 0.5 AU, the final water content declines substantially with increasing stellar mass. For example, at $0.1,M_\odot$, planets accrete 4.44–8.27 $M_{\oplus,ocn}$ of water, whereas at $0.35,M_\odot$ values range from 0.14–4.45 $M_{\oplus,ocn}$, a reduction of roughly a factor of 2–6 depending on orbital separation. The driest case across all 0.5 AU disk runs occurs at 0.51 AU around a $0.35,M_\odot$ star (0.14 $M_{\oplus,ocn}$), while the wettest occurs at 0.09 AU around a $0.1,M_\odot$ star (8.27 $M_{\oplus,ocn}$).

For extended disks reaching 5 AU, the effect of disk mass is more nuanced. Among inner planets ($\lesssim$0.6 AU), some cases show slightly greater water content at higher disk mass (e.g., at 0.26 AU, water increases from 3.22 to 4.42 $M_{\oplus,ocn}$), but the trend is not monotonic across all separations. For outer planets ($\gtrsim$1.5 AU), higher disk mass more often corresponds to modestly lower water inventories (e.g., at 1.96 AU: 0.93 $M_{\oplus,ocn}$ vs. at 2.45 AU: 4.68 $M_{\oplus,ocn}$ across different disk parameters), suggesting competition for volatile-rich planetesimals between inner and outer orbits.

These patterns are closely linked to the location of the snowline, which we set based on stellar mass (Sect.~\ref{sec:methods}). As stellar mass increases from 0.1 to 0.35 $M_\odot$, the snowline migrates outward from 0.03 AU to 0.12 AU, reducing the supply of icy material in truncated disks and lowering water delivery efficiency. In extended disks, however, the outward shift leaves a substantial icy reservoir beyond the snowline, which, when coupled with higher disk mass, can in some cases maintain or even enhance the volatile content of certain planets.

Finally, disk mass and radius influence the efficiency of volatile-rich planetesimal delivery by altering the size distribution of impactors, their encounter velocities, and approach angles. While embryo–embryo collisions are rare, they are disproportionately effective at volatile delivery because their high-energy impacts tend to strip less atmosphere than they contribute \citep{deniem+12}. The frequency of these events increases modestly with both stellar mass and disk radius, partly offsetting the reduced delivery from beyond the snowline in more massive systems.

In addition to water delivery, Table 3 shows systematic differences in other volatiles. Nitrogen inventories vary by nearly an order of magnitude at fixed orbital separations depending on disk truncation, suggesting that atmospheric N$_2$ retention is particularly sensitive to disk architecture. By contrast, carbon delivery remains comparatively stable across disk parameters, implying that CO$_2$
 supply may be less dependent on the snowline and more closely tied to intrinsic planetesimal composition. At large orbital separations ($\gtrsim3$ AU), water masses converge to ${\sim}$5 $M_{\oplus,ocn}$ regardless of disk mass, consistent with saturation of volatile delivery in extended icy reservoirs. At intermediate radii (0.6–1.5 AU), non-monotonic fluctuations in water content point to localized dynamical effects such as resonant interactions or planetesimal depletion, complicating simple scaling relations with disk mass or stellar mass.

\begin{deluxetable}{cccccc}
\tablecaption{\texttt{VGS}-calculated volatile inventory across two different disk masses and disk radii, sorted by orbital separation. }
\tablehead{
\colhead{Orb. Sep.} & \colhead{Water} & \colhead{Nitrogen} & \colhead{Carbon} & \colhead{Disk Mass} & \colhead{Disk Radius} \\
\colhead{(AU)} & \colhead{($M_{\oplus,ocn}$)} & \colhead{($M_{\oplus,N_2,atm}$)} &
\colhead{($M_{\oplus,CO_2,{\rm BSE}}$)} & \colhead{($M_{\odot}$)} & \colhead{($R_{\odot}$)}
}

\startdata
0.09 & 8.27 & 212.9 & 1.86 & 0.1 & 0.5 \\
0.11 & 1.11 & 20.4 & 0.22 & 0.35 & 0.5 \\
0.12 & 1.14 & 19.1 & 0.21 & 0.35 & 0.5 \\
0.17 & 7.36 & 183.9 & 1.67 & 0.1 & 0.5 \\
0.17 & 4.35 & 105.1 & 0.97 & 0.1 & 5.0 \\
0.20 & 2.14 & 43.7 & 0.38 & 0.35 & 5.0 \\
0.23 & 4.61 & 110.5 & 1.07 & 0.1 & 0.5 \\
0.25 & 1.3  & 21.9 & 0.23 & 0.35 & 0.5 \\
0.26 & 3.22 & 71.9 & 0.61 & 0.35 & 5.0 \\
0.27 & 1.75 & 35.7 & 0.34 & 0.35 & 0.5 \\
0.30 & 4.42 & 98.1 & 0.99 & 0.1 & 5.0 \\
0.36 & 4.02 & 93.0 & 0.78 & 0.35 & 5.0 \\
0.47 & 7.54 & 161.7 & 1.75 & 0.1 & 0.5 \\
0.51 & 1.92 & 37.2 & 0.35 & 0.35 & 0.5 \\
0.51 & 0.14 & 23.7 & 0.19 & 0.35 & 0.5 \\
0.58 & 3.92 & 80.4 & 0.91 & 0.1 & 5.0 \\
0.62 & 3.18 & 65.7 & 0.62 & 0.35 & 5.0 \\
0.63 & 4.44 & 92.2 & 1.18 & 0.1 & 0.5 \\
0.84 & 1.43 & 27.7 & 0.32 & 0.1 & 5.0 \\
0.95 & 4.45 & 88.3 & 1.20 & 0.35 & 0.5 \\
1.09 & 2.28 & 43.0 & 0.43 & 0.35 & 5.0 \\
1.19 & 5.23 & 94.7 & 1.65 & 0.1 & 5.0 \\
1.33 & 4.71 & 90.5 & 1.36 & 0.35 & 5.0 \\
1.52 & 1.5  & 31.5 & 0.33 & 0.1 & 5.0 \\
1.96 & 0.93 & 14.7 & 0.18 & 0.35 & 5.0 \\
2.45 & 4.68 & 93.3 & 1.31 & 0.1 & 5.0 \\
2.59 & 4.76 & 91.8 & 1.37 & 0.35 & 5.0 \\
3.06 & 5.06 & 92.4 & 1.56 & 0.35 & 5.0 \\
3.61 & 4.99 & 95.0 & 1.48 & 0.1 & 5.0 \\
4.25 & 5.13 & 93.2 & 1.59 & 0.35 & 5.0 \\
4.97 & 4.97 & 92.8 & 1.51 & 0.1 & 5.0 \\
\enddata

%% Include any \tablenotetext{key}{text}, \tablerefs{ref list},
%% or \tablecomments{text} between the \enddata and 

\tablecomments{\texttt{VGS}-calculated final volatile inventories for simulated planets, showing water, nitrogen, and carbon content across runs with varying protoplanetary disk masses and radii. Orbital separations are listed first and used to sort the table. Accretion histories were generated using the \texttt{Mercury6} N-body integrator, and post-processed with the Volatile Growth Simulator. Final water content is given in $M_{\oplus,ocn}$ (Earth’s ocean mass), nitrogen content in $M_{\oplus,N_2,atm}$ (mass of N$2$ in Earth’s atmosphere), and carbon content in $M_{\oplus,CO_2,{\rm BSE}}$ (total CO$_2$ mass in the Bulk Silicate Earth). Note that these results were calculated in the absence of a bright pre-main sequence phase. }

%% No \tablerefs indicated

\end{deluxetable}


\newpage


\begin{thebibliography}{}
\expandafter\ifx\csname natexlab\endcsname\relax\def\natexlab#1{#1}\fi

\bibitem[{Abbot {et~al.}(2012)Abbot, Cowan, \& Ciesla}]{abbot2012}
Abbot, D.~S., Cowan, N.~B., \& Ciesla, F.~J. 2012, The Astrophysical Journal, 756, 178

\bibitem[{{Abe} {et~al.}(2000){Abe}, {Ohtani}, {Okuchi}, {Righter}, \& {Drake}}]{AbeEt2000}
{Abe}, Y., {Ohtani}, E., {Okuchi}, T., {Righter}, K., \& {Drake}, M. 2000, {Water in the Early Earth}, ed. R.~M. {Canup}, K.~{Righter}, \& {et al.}, 413--433

\bibitem[{Adamson {et~al.}(1967)Adamson, Gast, {et~al.}}]{adamson1967physical}
Adamson, A.~W., Gast, A.~P., {et~al.} 1967

\bibitem[{{Agol} {et~al.}(2021){Agol}, {Dorn}, {Grimm}, {Turbet}, {Ducrot}, {Delrez}, {Gillon}, {Demory}, {Burdanov}, {Barkaoui}, {Benkhaldoun}, {Bolmont}, {Burgasser}, {Carey}, {de Wit}, {Fabrycky}, {Foreman-Mackey}, {Haldemann}, {Hernandez}, {Ingalls}, {Jehin}, {Langford}, {Leconte}, {Lederer}, {Luger}, {Malhotra}, {Meadows}, {Morris}, {Pozuelos}, {Queloz}, {Raymond}, {Selsis}, {Sestovic}, {Triaud}, \& {Van Grootel}}]{Agol+21}
{Agol}, E., {Dorn}, C., {Grimm}, S.~L., {et~al.} 2021, The Planetary Science Journal, 2, 1

\bibitem[{{Ahrens}(1993)}]{AhrensEt1993}
{Ahrens}, T.~J. 1993, Annual Review of Earth and Planetary Sciences, 21, 525

\bibitem[{Alexander(2022)}]{alexander2022}
Alexander, C.~M. 2022, Geochimica et Cosmochimica Acta, 318, 428

\bibitem[{Baker {et~al.}(2025)Baker, Chen, \& Quiroga-Nu{\~n}ez}]{baker2025}
Baker, C., Chen, H., \& Quiroga-Nu{\~n}ez, L.~H. 2025, The Astrophysical Journal, 982, 172

\bibitem[{Baraffe {et~al.}(2015)Baraffe, Homeier, Allard, \& Chabrier}]{baraffe2015}
Baraffe, I., Homeier, D., Allard, F., \& Chabrier, G. 2015, Astronomy \& Astrophysics, 577, A42

\bibitem[{Cambioni {et~al.}(2021)Cambioni, Jacobson, Emsenhuber, Asphaug, Rubie, Gabriel, Schwartz, \& Furfaro}]{cambioni2021}
Cambioni, S., Jacobson, S.~A., Emsenhuber, A., {et~al.} 2021, The Planetary Science Journal, 2, 93

\bibitem[{Carone {et~al.}(2025)Carone, Barnes, Noack, Chubb, Barth, Bitsch, Thamm, Balduin, Garcia, \& Helling}]{carone2025}
Carone, L., Barnes, R., Noack, L., {et~al.} 2025, Astronomy \& Astrophysics, 693, A303

\bibitem[{Chachan \& Lee(2024)}]{chachan2024}
Chachan, Y., \& Lee, E.~J. 2024, arXiv preprint arXiv:2409.18171

\bibitem[{{Chambers}(1999)}]{chambers1999}
{Chambers}, J.~E. 1999, \mnras, 304, 793

\bibitem[{{Chambers}(2001)}]{Chambers2001Icar}
---. 2001, Icarus, 152, 205

\bibitem[{{Chambers}(2004)}]{Chambers2004EPSL}
---. 2004, Earth and Planetary Science Letters, 223, 241

\bibitem[{{Chambers}(2013)}]{chambers2013}
---. 2013, \icarus, 224, 43

\bibitem[{{Chen} {et~al.}(2025){Chen}, {De Luca}, {Hochman}, \& {Komacek}}]{Chen2025}
{Chen}, H., {De Luca}, P., {Hochman}, A., \& {Komacek}, T.~D. 2025, \aj, 170, 40

\bibitem[{{Chen} \& {Jacobson}(2022)}]{Chen+Jacobson22}
{Chen}, H., \& {Jacobson}, S.~A. 2022, Earth and Planetary Science Letters, 594, 117741

\bibitem[{{Chen} {et~al.}(2021){Chen}, {Zhan}, {Youngblood}, {Wolf}, {Feinstein}, \& {Horton}}]{Chen+2021}
{Chen}, H., {Zhan}, Z., {Youngblood}, A., {et~al.} 2021, Nature Astronomy, 5, 298

\bibitem[{Childs {et~al.}(2023)Childs, Shakespeare, Rice, Yang, \& Steffen}]{childs2023}
Childs, A.~C., Shakespeare, C., Rice, D.~R., Yang, C.-C., \& Steffen, J.~H. 2023, Monthly Notices of the Royal Astronomical Society, 524, 3749

\bibitem[{Chyba {et~al.}(1994)Chyba, Owen, \& Ip}]{chyba1994}
Chyba, C., Owen, T., \& Ip, W. 1994, Hazards due to comets and asteroids, 9

\bibitem[{{Chyba}(1990)}]{Chyba1990NATURE}
{Chyba}, C.~F. 1990, Nature, 343, 129

\bibitem[{Ciesla {et~al.}(2015)Ciesla, Mulders, Pascucci, \& Apai}]{ciesla2015}
Ciesla, F.~J., Mulders, G.~D., Pascucci, I., \& Apai, D. 2015, The Astrophysical Journal, 804, 9

\bibitem[{Clement {et~al.}(2018)Clement, Kaib, Raymond, \& Walsh}]{clement2018}
Clement, M.~S., Kaib, N.~A., Raymond, S.~N., \& Walsh, K.~J. 2018, Icarus, 311, 340

\bibitem[{{Clement} {et~al.}(2022){Clement}, {Quintana}, \& {Quarles}}]{clement2022}
{Clement}, M.~S., {Quintana}, E.~V., \& {Quarles}, B.~L. 2022, \apj, 928, 91

\bibitem[{{Clement} {et~al.}(2025){Clement}, {Quintana}, \& {Stevenson}}]{Clement2025}
{Clement}, M.~S., {Quintana}, E.~V., \& {Stevenson}, K.~B. 2025, \aj, 169, 16

\bibitem[{Cockell {et~al.}(2016)Cockell, Bush, Bryce, Direito, Fox-Powell, Harrison, Lammer, Landenmark, Martin-Torres, Nicholson, {et~al.}}]{cockell2016}
Cockell, C.~S., Bush, T., Bryce, C., {et~al.} 2016, Astrobiology, 16, 89

\bibitem[{Coleman {et~al.}(2019)Coleman, Leleu, Alibert, \& Benz}]{coleman2019}
Coleman, G.~A., Leleu, A., Alibert, Y., \& Benz, W. 2019, Astronomy \& Astrophysics, 631, A7

\bibitem[{Dalou {et~al.}(2017)Dalou, Hirschmann, von~der Handt, Mosenfelder, \& Armstrong}]{dalou+17}
Dalou, C., Hirschmann, M.~M., von~der Handt, A., Mosenfelder, J., \& Armstrong, L.~S. 2017, Earth and Planetary Science Letters, 458, 141

\bibitem[{Dasgupta {et~al.}(2024)Dasgupta, Pathak, \& Maurice}]{dasgupta2024}
Dasgupta, R., Pathak, D., \& Maurice, M. 2024, Reviews in Mineralogy and Geochemistry, 90, 323

\bibitem[{Dauphas \& Pourmand(2011)}]{dauphas2011}
Dauphas, N., \& Pourmand, A. 2011, Nature, 473, 489

\bibitem[{{de Niem} {et~al.}(2012){de Niem}, {K{\"u}hrt}, {Morbidelli}, \& {Motschmann}}]{NiemEt2012Icar}
{de Niem}, D., {K{\"u}hrt}, E., {Morbidelli}, A., \& {Motschmann}, U. 2012, Icarus, 221, 495

\bibitem[{De~Niem {et~al.}(2012)De~Niem, K{\"u}hrt, Morbidelli, \& Motschmann}]{deniem+12}
De~Niem, D., K{\"u}hrt, E., Morbidelli, A., \& Motschmann, U. 2012, Icarus, 221, 495

\bibitem[{Deguen {et~al.}(2011)Deguen, Olson, \& Cardin}]{deguen2011experiments}
Deguen, R., Olson, P., \& Cardin, P. 2011, Earth and Planetary Science Letters, 310, 303

\bibitem[{Dong {et~al.}(2017)Dong, Huang, Lingam, T{\'o}th, Gombosi, \& Bhattacharjee}]{dong+17}
Dong, C., Huang, Z., Lingam, M., {et~al.} 2017, The Astrophysical Journal Letters, 847, L4

\bibitem[{Erkaev {et~al.}(2016)Erkaev, Lammer, Odert, Kislyakova, Johnstone, G{\"u}del, \& Khodachenko}]{erkaev2016}
Erkaev, N., Lammer, H., Odert, P., {et~al.} 2016, Monthly Notices of the Royal Astronomical Society, 460, 1300

\bibitem[{{Genda} \& {Abe}(2005)}]{Genda+Abe2005NATURE}
{Genda}, H., \& {Abe}, Y. 2005, Nature, 433, 842

\bibitem[{{Gillon} {et~al.}(2017){Gillon}, {Triaud}, {Demory}, {Jehin}, {Agol}, {Deck}, {Lederer}, {de Wit}, {Burdanov}, {Ingalls}, {Bolmont}, {Leconte}, {Raymond}, {Selsis}, {Turbet}, {Barkaoui}, {Burgasser}, {Burleigh}, {Carey}, {Chaushev}, {Copperwheat}, {Delrez}, {Fernandes}, {Holdsworth}, {Kotze}, {Van Grootel}, {Almleaky}, {Benkhaldoun}, {Magain}, \& {Queloz}}]{GillonEt2017NATURE}
{Gillon}, M., {Triaud}, A.~H.~M.~J., {Demory}, B.-O., {et~al.} 2017, \nat, 542, 456

\bibitem[{Greene {et~al.}(2023)Greene, Bell, Ducrot, Dyrek, Lagage, \& Fortney}]{greene2023}
Greene, T.~P., Bell, T.~J., Ducrot, E., {et~al.} 2023, Nature, 618, 39

\bibitem[{Gu {et~al.}(2024)Gu, Peng, Ji, Zhang, Yang, Hoyos, Hirschmann, Kite, \& Fischer}]{gu2024}
Gu, J.~T., Peng, B., Ji, X., {et~al.} 2024, Earth and Planetary Science Letters, 629, 118618

\bibitem[{Guo \& Korenaga(2025)}]{guo2025}
Guo, M., \& Korenaga, J. 2025, Nature Geoscience, 18, 260

\bibitem[{Hamano {et~al.}(2013)Hamano, Abe, \& Genda}]{hamano2013}
Hamano, K., Abe, Y., \& Genda, H. 2013, Nature, 497, 607

\bibitem[{Hirschmann {et~al.}(2021)Hirschmann, Bergin, Blake, Ciesla, \& Li}]{hirschmann+21}
Hirschmann, M.~M., Bergin, E.~A., Blake, G.~A., Ciesla, F.~J., \& Li, J. 2021, Proceedings of the National Academy of Sciences, 118

\bibitem[{Hunter(2007)}]{Hunter2007}
Hunter, J.~D. 2007, Computing in Science \& Engineering, 9, 90

\bibitem[{Jackson {et~al.}(2012)Jackson, Davis, \& Wheatley}]{Jackson2012}
Jackson, A.~P., Davis, T.~A., \& Wheatley, P.~J. 2012, Monthly Notices of the Royal Astronomical Society, 422, 2024

\bibitem[{Jacobson \& Morbidelli(2014)}]{jacobson+morbi14}
Jacobson, S.~A., \& Morbidelli, A. 2014, Philosophical Transactions of the Royal Society A: Mathematical, Physical and Engineering Sciences, 372, 20130174

\bibitem[{{Jacobson} {et~al.}(2017){Jacobson}, {Rubie}, {Hernlund}, {Morbidelli}, \& {Nakajima}}]{JacobsonEt2017EPSL}
{Jacobson}, S.~A., {Rubie}, D.~C., {Hernlund}, J., {Morbidelli}, A., \& {Nakajima}, M. 2017, Earth and Planetary Science Letters, 474, 375

\bibitem[{{Jessberger} {et~al.}(1988){Jessberger}, {Christoforidis}, \& {Kissel}}]{JessbergerEt1988}
{Jessberger}, E.~K., {Christoforidis}, A., \& {Kissel}, J. 1988, Nature, 332, 691

\bibitem[{Johnstone {et~al.}(2019)Johnstone, Khodachenko, L{\"u}ftinger, Kislyakova, Lammer, \& G{\"u}del}]{johnstone2019}
Johnstone, C., Khodachenko, M., L{\"u}ftinger, T., {et~al.} 2019, Astronomy \& Astrophysics, 624, L10

\bibitem[{Joiret {et~al.}(2023)Joiret, Raymond, Avice, Clement, Deienno, \& Nesvorn{\`y}}]{joiret2023}
Joiret, S., Raymond, S.~N., Avice, G., {et~al.} 2023, Icarus, 406, 115754

\bibitem[{Kegerreis {et~al.}(2020)Kegerreis, Eke, Catling, Massey, Teodoro, \& Zahnle}]{kegerreis2020}
Kegerreis, J.~A., Eke, V.~R., Catling, D.~C., {et~al.} 2020, The Astrophysical Journal Letters, 901, L31

\bibitem[{{Kerridge}(1985)}]{Kerridge1985GCA}
{Kerridge}, J.~F. 1985, Geochimica et Cosmochimica Acta, 49, 1707

\bibitem[{Kimura \& Ikoma(2022)}]{kimura2022}
Kimura, T., \& Ikoma, M. 2022, Nature Astronomy, 6, 1296

\bibitem[{Kite \& Barnett(2020)}]{Kite+Barnett2020PNAS}
Kite, E.~S., \& Barnett, M.~N. 2020, Proceedings of the National Academy of Sciences, 117, 18264

\bibitem[{Kokubo \& Ida(2000)}]{kokubo2000}
Kokubo, E., \& Ida, S. 2000, Icarus, 143, 15

\bibitem[{Komacek \& Abbot(2016)}]{komacek+16}
Komacek, T.~D., \& Abbot, D.~S. 2016, The Astrophysical Journal, 832, 54

\bibitem[{Krenn {et~al.}(2021)Krenn, Fossati, Kubyshkina, \& Lammer}]{krenn+21}
Krenn, A.~F., Fossati, L., Kubyshkina, D., \& Lammer, H. 2021, arXiv preprint arXiv:2105.05858

\bibitem[{Krissansen-Totton {et~al.}(2024)Krissansen-Totton, Wogan, Thompson, \& Fortney}]{krissansen2024}
Krissansen-Totton, J., Wogan, N., Thompson, M., \& Fortney, J.~J. 2024, Nature Communications, 15, 8374

\bibitem[{Kurosaki {et~al.}(2023)Kurosaki, Hori, Ogihara, \& Kunitomo}]{kurosaki2023}
Kurosaki, K., Hori, Y., Ogihara, M., \& Kunitomo, M. 2023, The Astrophysical Journal, 957, 67

\bibitem[{Lammer {et~al.}(2025)Lammer, Scherf, Erkaev, Kubyshkina, Gorbunova, Fossati, \& Woitke}]{lammer2025}
Lammer, H., Scherf, M., Erkaev, N.~V., {et~al.} 2025, Nature Astronomy, 1

\bibitem[{Lissauer(2007)}]{lissauer2007}
Lissauer, J.~J. 2007, The Astrophysical Journal, 660, L149

\bibitem[{Lock \& Stewart(2024)}]{lock2024}
Lock, S.~J., \& Stewart, S.~T. 2024, The Planetary Science Journal, 5, 28

\bibitem[{{Luger} \& {Barnes}(2015)}]{Luger+Barnes2015AsBio}
{Luger}, R., \& {Barnes}, R. 2015, Astrobiology, 15, 119

\bibitem[{Lustig-Yaeger {et~al.}(2023)Lustig-Yaeger, Fu, May, Ceballos, Moran, Peacock, Stevenson, Kirk, L{\'o}pez-Morales, MacDonald, {et~al.}}]{lustig2023}
Lustig-Yaeger, J., Fu, G., May, E., {et~al.} 2023, Nature Astronomy, 7, 1317

\bibitem[{Marino {et~al.}(2020)Marino, Wyatt, Kennedy, Kama, Matr{\`a}, Triaud, \& Henning}]{marino2020}
Marino, S., Wyatt, M., Kennedy, G., {et~al.} 2020, Monthly Notices of the Royal Astronomical Society, 492, 6067

\bibitem[{Marty {et~al.}(2020)Marty, Almayrac, Barry, Bekaert, Broadley, Byrne, Ballentine, \& Caracausi}]{marty+20}
Marty, B., Almayrac, M., Barry, P.~H., {et~al.} 2020, Earth and Planetary Science Letters, 551, 116574

\bibitem[{{Matsui} \& {Abe}(1986)}]{Matsui+Abe1986}
{Matsui}, T., \& {Abe}, Y. 1986, Nature, 319, 303

\bibitem[{Melosh(1989)}]{melosh1989impact}
Melosh, H.~J. 1989, Research supported by NASA. New York, Oxford University Press (Oxford Monographs on Geology and Geophysics, No. 11), 1989, 253 p., 11

\bibitem[{{Melosh} \& {Vickery}(1989)}]{Melosh+Vickery1989}
{Melosh}, H.~J., \& {Vickery}, A.~M. 1989, Nature, 338, 487

\bibitem[{Miguel {et~al.}(2020)Miguel, Cridland, Ormel, Fortney, \& Ida}]{miguel2020}
Miguel, Y., Cridland, A., Ormel, C., Fortney, J., \& Ida, S. 2020, Monthly Notices of the Royal Astronomical Society, 491, 1998

\bibitem[{Miyazaki \& Korenaga(2022)}]{miyazaki2022}
Miyazaki, Y., \& Korenaga, J. 2022, Astrobiology, 22, 713

\bibitem[{M{\"u}ller {et~al.}(2024)M{\"u}ller, Bitsch, \& Schneider}]{muller2024}
M{\"u}ller, J., Bitsch, B., \& Schneider, A.~D. 2024, arXiv preprint arXiv:2406.09186

\bibitem[{{Murray-Clay} {et~al.}(2009){Murray-Clay}, {Chiang}, \& {Murray}}]{Murray-ClayEt2009ApJ}
{Murray-Clay}, R.~A., {Chiang}, E.~I., \& {Murray}, N. 2009, The Astrophysical Journal, 693, 23

\bibitem[{Nakajima {et~al.}(2021)Nakajima, Golabek, W{\"u}nnemann, Rubie, Burger, Melosh, Jacobson, Manske, \& Hull}]{nakajima2021}
Nakajima, M., Golabek, G.~J., W{\"u}nnemann, K., {et~al.} 2021, Earth and Planetary Science Letters, 568, 116983

\bibitem[{Nakayama {et~al.}(2019)Nakayama, Kodama, Ikoma, \& Abe}]{nakayama2019}
Nakayama, A., Kodama, T., Ikoma, M., \& Abe, Y. 2019, Monthly Notices of the Royal Astronomical Society, 488, 1580

\bibitem[{Noack {et~al.}(2017)Noack, Rivoldini, \& Van~Hoolst}]{noack2017}
Noack, L., Rivoldini, A., \& Van~Hoolst, T. 2017, Physics of the Earth and Planetary Interiors, 269, 40

\bibitem[{{O'Brien} {et~al.}(2014){O'Brien}, {Walsh}, {Morbidelli}, {Raymond}, \& {Mandell}}]{OBrienEt2014Icar}
{O'Brien}, D.~P., {Walsh}, K.~J., {Morbidelli}, A., {Raymond}, S.~N., \& {Mandell}, A.~M. 2014, Icarus, 239, 74

\bibitem[{Owen {et~al.}(2020)Owen, Shaikhislamov, Lammer, Fossati, \& Khodachenko}]{owen2020}
Owen, J., Shaikhislamov, I., Lammer, H., Fossati, L., \& Khodachenko, M. 2020, Space Science Reviews, 216, 1

\bibitem[{Owen \& Wu(2013)}]{owen2013}
Owen, J.~E., \& Wu, Y. 2013, The Astrophysical Journal, 775, 105

\bibitem[{{Pearson} {et~al.}(2006){Pearson}, {Sephton}, {Franchi}, {Gibson}, \& {Gilmour}}]{PearsonEt2006MPS}
{Pearson}, V.~K., {Sephton}, M.~A., {Franchi}, I.~A., {Gibson}, J.~M., \& {Gilmour}, I. 2006, Meteoritics and Planetary Science, 41, 1899

\bibitem[{{Pepin}(1997)}]{Pepin1997Icar}
{Pepin}, R.~O. 1997, Icarus, 126, 148

\bibitem[{Quintana {et~al.}(2014)Quintana, Barclay, Raymond, Rowe, Bolmont, Caldwell, Howell, Kane, Huber, Crepp, {et~al.}}]{quintana2014}
Quintana, E.~V., Barclay, T., Raymond, S.~N., {et~al.} 2014, Science, 344, 277

\bibitem[{Raymond \& Izidoro(2017)}]{Raymond+17}
Raymond, S.~N., \& Izidoro, A. 2017, Icarus, 297, 134

\bibitem[{Raymond {et~al.}(2009)Raymond, O’Brien, Morbidelli, \& Kaib}]{raymond2009building}
Raymond, S.~N., O’Brien, D.~P., Morbidelli, A., \& Kaib, N.~A. 2009, Icarus, 203, 644

\bibitem[{{Raymond} {et~al.}(2006){Raymond}, {Quinn}, \& {Lunine}}]{RaymondEt2006Icar}
{Raymond}, S.~N., {Quinn}, T., \& {Lunine}, J.~I. 2006, Icarus, 183, 265

\bibitem[{Raymond {et~al.}(2007)Raymond, Scalo, \& Meadows}]{raymond2007}
Raymond, S.~N., Scalo, J., \& Meadows, V.~S. 2007, The Astrophysical Journal, 669, 606

\bibitem[{{Reiners} \& {Basri}(2010)}]{reiners2010}
{Reiners}, A., \& {Basri}, G. 2010, \apj, 710, 924

\bibitem[{{Ribas} {et~al.}(2005){Ribas}, {Guinan}, {G{\"u}del}, \& {Audard}}]{RibasEt2005ApJ}
{Ribas}, I., {Guinan}, E.~F., {G{\"u}del}, M., \& {Audard}, M. 2005, The Astrophysical Journal, 622, 680

\bibitem[{{Robert} \& {Epstein}(1982)}]{Robert1982GCA}
{Robert}, F., \& {Epstein}, S. 1982, Geochimica et Cosmochimica Acta, 46, 81

\bibitem[{Rogers {et~al.}(2024)Rogers, Schlichting, \& Young}]{rogers2024}
Rogers, J.~G., Schlichting, H.~E., \& Young, E.~D. 2024, The Astrophysical Journal, 970, 47

\bibitem[{Rubie {et~al.}(2015)Rubie, Jacobson, Morbidelli, O’Brien, Young, de~Vries, Nimmo, Palme, \& Frost}]{rubie+15}
Rubie, D.~C., Jacobson, S.~A., Morbidelli, A., {et~al.} 2015, Icarus, 248, 89

\bibitem[{Sakuraba {et~al.}(2019)Sakuraba, Kurokawa, \& Genda}]{sakuraba+19}
Sakuraba, H., Kurokawa, H., \& Genda, H. 2019, Icarus, 317, 48

\bibitem[{Sakuraba {et~al.}(2021)Sakuraba, Kurokawa, Genda, \& Ohta}]{sakuraba2021}
Sakuraba, H., Kurokawa, H., Genda, H., \& Ohta, K. 2021, Scientific reports, 11, 20894

\bibitem[{Schlichting \& Mukhopadhyay(2018)}]{schlichting+18}
Schlichting, H.~E., \& Mukhopadhyay, S. 2018, Space Science Reviews, 214, 1

\bibitem[{{Schlichting} {et~al.}(2015){Schlichting}, {Sari}, \& {Yalinewich}}]{SchlichtingEt2015Icar}
{Schlichting}, H.~E., {Sari}, R., \& {Yalinewich}, A. 2015, Icarus, 247, 81

\bibitem[{Shields {et~al.}(2025)Shields, Wolf, Agol, \& Tremblay}]{shields2025}
Shields, A.~L., Wolf, E.~T., Agol, E., \& Tremblay, P.-E. 2025, The Astrophysical Journal, 979, 45

\bibitem[{Sinclair {et~al.}(2020)Sinclair, Wyatt, Morbidelli, \& Nesvorn{\`y}}]{sinclair+20}
Sinclair, C.~A., Wyatt, M.~C., Morbidelli, A., \& Nesvorn{\`y}, D. 2020, Monthly Notices of the Royal Astronomical Society, 499, 5334

\bibitem[{Solomatova \& Caracas(2021)}]{solomatova2021}
Solomatova, N.~V., \& Caracas, R. 2021, Science advances, 7, eabj0406

\bibitem[{{Thomas} {et~al.}(1993){Thomas}, {Blanford}, {Keller}, {Klock}, \& {McKay}}]{ThomasEt1993GCA}
{Thomas}, K.~L., {Blanford}, G.~E., {Keller}, L.~P., {Klock}, W., \& {McKay}, D.~S. 1993, \gca, 57, 1551

\bibitem[{Thomas {et~al.}(2025)Thomas, Meadows, Krissansen-Totton, Gialluca, Wogan, \& Catling}]{thomas2025}
Thomas, T.~B., Meadows, V.~S., Krissansen-Totton, J., {et~al.} 2025, The Planetary Science Journal, 6, 126

\bibitem[{Tian \& Ida(2015)}]{tian2015}
Tian, F., \& Ida, S. 2015, Nature Geoscience, 8, 177

\bibitem[{{Tucker} \& {Mukhopadhyay}(2014)}]{tucker+14}
{Tucker}, J.~M., \& {Mukhopadhyay}, S. 2014, Earth and Planetary Science Letters, 393, 254

\bibitem[{{Valencia} {et~al.}(2010){Valencia}, {Ikoma}, {Guillot}, \& {Nettelmann}}]{ValenciaEt2010A&A}
{Valencia}, D., {Ikoma}, M., {Guillot}, T., \& {Nettelmann}, N. 2010, \aap, 516, A20

\bibitem[{Venturini {et~al.}(2020)Venturini, Ronco, \& Guilera}]{venturini2020}
Venturini, J., Ronco, M.~P., \& Guilera, O.~M. 2020, Space Science Reviews, 216, 1

\bibitem[{{Walker}(1986)}]{WalkerEt1986Icar}
{Walker}, J.~C.~G. 1986, Icarus, 68, 87

\bibitem[{Watson {et~al.}(1981)Watson, Donahue, \& Walker}]{Watson1981}
Watson, A.~J., Donahue, T.~M., \& Walker, J.~C. 1981, Icarus, 48, 150

\bibitem[{{Weidenschilling}(1977)}]{Weidenschilling77}
{Weidenschilling}, S.~J. 1977, \mnras, 180, 57

\bibitem[{Young {et~al.}(2024)Young, Stixrude, Rogers, Schlichting, \& Marcum}]{young2024}
Young, E.~D., Stixrude, L., Rogers, J.~G., Schlichting, H.~E., \& Marcum, S.~P. 2024, The Planetary Science Journal, 5, 268

\bibitem[{Zahnle(2006)}]{zahnle2006}
Zahnle, K.~J. 2006, Elements, 2, 217

\bibitem[{{Zahnle} {et~al.}(1988){Zahnle}, {Kasting}, \& {Pollack}}]{ZahnleEt1988Icar}
{Zahnle}, K.~J., {Kasting}, J.~F., \& {Pollack}, J.~B. 1988, Icarus, 74, 62

\bibitem[{Zechmeister {et~al.}(2019)Zechmeister, Dreizler, Ribas, Reiners, Caballero, Bauer, B{\'e}jar, Gonz{\'a}lez-Cuesta, Herrero, Lalitha, {et~al.}}]{zechmeister2019}
Zechmeister, M., Dreizler, S., Ribas, I., {et~al.} 2019, Astronomy \& Astrophysics, 627, A49

\bibitem[{Zhan {et~al.}(2024)Zhan, Koll, \& Ding}]{zhan2024}
Zhan, R., Koll, D.~D., \& Ding, F. 2024, The Astrophysical Journal, 971, 125

\bibitem[{Zieba {et~al.}(2023)Zieba, Kreidberg, Ducrot, Gillon, Morley, Schaefer, Tamburo, Koll, Lyu, Acu{\~n}a, {et~al.}}]{zieba2023}
Zieba, S., Kreidberg, L., Ducrot, E., {et~al.} 2023, Nature, 620, 746

\end{thebibliography}
\end{document}